\documentclass[aps,onecolumn,floatfix,superscriptaddress,longbibliography,notitlepage, nofootinbib]{revtex4-2}
\pdfoutput=1

\usepackage[utf8]{inputenc}
\usepackage{braket}
\usepackage{amsmath}
\DeclareMathOperator{\Tr}{Tr}
\usepackage{dcolumn}   % needed for some tables
\usepackage{bm}        % for math
\usepackage{amssymb}
\usepackage{mathtools}
\usepackage{tikz}
\usepackage{pgfplots}
\usepackage{amsfonts}
\usepackage{esvect}
\usepackage{mathrsfs}

\usepackage{appendix}
\usepackage{hyperref}
\usepackage[font=small,labelfont=bf, justification=justified, format=plain]{caption}

\usepackage{mathtools}
\usepackage{thmtools,thm-restate}
\usepackage{algorithm}

\usepackage{algpseudocode}

\newtheorem{definition}{Definition}
\newtheorem{theorem}{Theorem}
\newtheorem{lemma}{Lemma}

\newtheorem{method}{Algorithm}

\newcommand{\revise}[1]{\textcolor{black}{#1}}
\newcommand{\xbf}{\textbf{x}}

\newcommand{\Ibb}{\mathbb{I}}

\usetikzlibrary{arrows.meta}

\def\BibTeX{{\rm B\kern-.05em{\sc i\kern-.025em b}\kern-.08em
    T\kern-.1667em\lower.7ex\hbox{E}\kern-.125emX}}

\begin{document}
\title{Improved Quantum Algorithms  for Eigenvalues Finding and Gradient Descent }
\author{Nhat A. Nghiem }
\email{nhatanh.nghiemvu@stonybrook.edu}
\affiliation{Department of Physics and Astronomy, State University of New York at Stony Brook, Stony Brook, NY 11794-3800, USA}
\author{Tzu-Chieh Wei}
\affiliation{Department of Physics and Astronomy, State University of New York at Stony Brook, Stony Brook, NY 11794-3800, USA}
\affiliation{C. N. Yang Institute for Theoretical Physics, State University of New York at Stony Brook, Stony Brook, NY 11794-3840, USA}
\begin{abstract}
Block encoding is a key ingredient in the recently developed \revise{quantum singular value transformation (QSVT) framework, which provides a unifying description for many quantum algorithms. Initially introduced to simplify and optimize resource utilization in various problems—such as searching, amplitude estimation, and Hamiltonian simulation.} \revise{It is reasonable to expect that the capabilities of QSVT} extend beyond these applications and offer untapped potential for designing new quantum algorithms.
\revise{In this article, we affirm this perspective} by leveraging block encoding to substantially enhance two previously proposed quantum algorithms: largest eigenvalue estimation and quantum gradient descent. Unlike previous works that rely on sophisticated approaches, our findings demonstrate that even just elementary operations within the unitary block encoding framework can eliminate major scaling factors present in their original counterparts. This results in significantly more efficient quantum algorithms capable of tackling target computational problems with remarkable efficiency.
Furthermore, we illustrate how our proposed method can be extended to other contexts, including matrix inversion and multiple eigenvalue estimation.

%Block encoding is a key ingredient in the recently developed quantum \revise{singular value transformation framework} that forms a unifying \revise{description} for many quantum algorithms. Initially showcased for simplifying and optimizing resource utilization in several problems, such as searching, amplitude estimation, and Hamiltonian simulation, it is \revise{reasonable to believe} that the capabilities of the quantum quantum singular value transformation go beyond these and likely offer untapped potential for devising new quantum algorithms. In this article, we \revise{affirm such a perspective by utilizing} block encoding to substantially enhance two previously proposed quantum algorithms: largest eigenvalue estimation and quantum gradient descent. Unlike previous works that involve sophisticated procedures, our findings, using the unitary block encoding, demonstrate that even with elementary operations, these newly revamped algorithms can shed major scaling factors present in their original counterparts. This yields much more efficient quantum algorithms capable of tackling complex computational problems with remarkable efficiency. Furthermore, we show how to extend our proposed method to different contexts, including matrix inversion and multiple eigenvalue estimation. 

\end{abstract}
\maketitle

\section{Introduction}
Quantum computing is currently experiencing rapid and exciting advancements due to its immense potential to tackle complex computational problems that often elude classical computing capability, as well as tremendous progress in quantum devices. There has \revise{been the} development of many quantum algorithms, a few of them exhibiting substantial speedup against known classical ones. For example, Grover's search algorithm~\cite{grover1996fast} demonstrated a quadratic speedup compared to the most efficient classical counterpart. On the other hand, Shor's algorithm~\cite{shor1999polynomial} and a recent improvement by Regev~\cite{regev2023efficient} showcase superpolynomial speedup in integer factorization. \revise{In particular}, a series of pivotal contributions~\cite{feynman2018simulating, lloyd1996universal, berry2007efficient, berry2012black, berry2015hamiltonian} yielded quantum algorithms \revise{to simulate} the dynamics of quantum systems,  which is also a critical and highly impactful application of quantum computing in the physics domain. Moreover, some recent works have demonstrated the potential of quantum computing in machine learning and artificial intelligence~\cite{lloyd2013quantum, lloyd2014quantum, schuld2014quest, schuld2018supervised, schuld2020circuit, schuld2019machine, havlivcek2019supervised}, which is arguably one of the \revise{main} focuses in industry.  

Among many quantum algorithmic breakthroughs, the Harrow-Hassidim-Lloyd (HHL) algorithm~\cite{harrow2009quantum} provides an exponential speedup for solving linear systems. This landmark achievement has opened up new horizons in quantum computing, given the foundational role that linear systems play across a wide array of engineering, theoretical, and applied scientific domains. In fact, various subsequent quantum algorithms have been built upon the framework of the HHL algorithm to tackle diverse computational challenges. Notable examples include quantum data fitting~\cite{wiebe2014quantum}, accurate estimation of electromagnetic cross sections~\cite{clader2013preconditioned}, creation of support vector machines for classification tasks~\cite{rebentrost2019quantum}, and solving linear and nonlinear partial differential equations~\cite{arrazola2019quantum, berry2014high, childs2021high}.

As the potential of quantum computers continues to unfold, exploring domains where quantum advantage can make a transformative impact is both natural and increasingly imperative. This quest to enhance the frontiers of computing is driven by the growing demand to harness the power of quantum advancements and unlock novel possibilities.

In the above, we have described various quantum algorithms, each tailored to address specific computational challenges. These algorithms exhibit unique execution patterns, aligning with their distinct problem-solving objectives. \revise{However, recent groundbreaking advances have shed new light on the field. Known as ``quantum singular value transformation''~\cite{gilyen2019quantum, martyn2021grand} (QSVT), this innovative framework has revealed a {\it unifying approach } that transcends individual differences among various quantum algorithms}. It should be noted that this concept can also be linked to ``quantum signal processing''~\cite{low2017optimal}, as a nearly simultaneous development occurred when the authors of~\cite{low2017optimal} proposed a simple yet remarkably powerful technique. Essentially, the underlying power of such a framework features two things. First, one encodes some matrix of interest into a unitary, i.e., the so-called unitary block encoding (see definition~\ref{def: blockencode} below). Second, one utilizes single and multi-qubit gates (e.g., the Toffoli gate) to transform such a block-encoded matrix to our desire. This approach, for instance, has constructed provably optimal quantum simulation algorithms~\cite{low2017optimal, low2019hamiltonian}. \revise{Over time, the QSVT framework has undergone multifaceted extensions, including the works of~\cite{tan2023error, mitarai2023perturbation, rossi2022multivariable, rall2020quantum, chakraborty2018power}, and underscores the dynamic nature of quantum algorithmic research and its ability to have a vast impact.}

\revise{As pointed out in the original construction \cite{gilyen2019quantum}, QSVT provides a unified description of many prior quantum algorithms and even simplifies the execution of multiple ones. A key and meaningful question arises: Can quantum computation be enhanced by quantum singular value transformation, or is it merely a new language that serves as a convenient tool for executing quantum algorithms? For certain quantum algorithms with established optimality, such as Grover's search algorithm, it is straightforward to deduce that no further asymptotic improvement is possible. However, on a more positive note, the work~\cite{gilyen2022quantum} has shown that QSVT enables a systematic and efficient method for building the Petz recovery channel, an important tool in quantum information science, while without QSVT, such a construction would be nearly impossible. Additionally, reference \cite{chakraborty_et_al:LIPIcs.ICALP.2019.33} showed that QSVT can significantly enhance regression techniques, yielding much lower complexity scaling (a polynomial speed-up). These examples highlight the remarkable power of QSVT and have motivated us to further expand the capabilities of the unitary block encoding method and, more broadly, the quantum singular value transformation framework.   }

\revise{
%We remark that while prior constructions, such as \cite{chakraborty_et_al:LIPIcs.ICALP.2019.33}, provide improved quantum algorithms by enhancing certain subroutines within the algorithm itself, we shall see that by adopting the perspective of block-encoded operators and related arithmetic techniques, the overall structure of quantum algorithms can also be transformed. As a result, more efficient algorithms can be achieved. Thus, we envision that QSVT not only unifies the description of quantum algorithms but also potentially offers a completely new paradigm for designing novel quantum algorithms. This work exemplifies this vision by leveraging unitary block encoding and related techniques to significantly enhance two previously proposed quantum algorithms.
\textit{In this work, we provide two examples of how the use of QSVT enhances their efficiency.} The first algorithm concerns the estimation of the largest eigenvalue of a given matrix $A$ through an oracle access~\cite{nghiem2022quantum}. The second algorithm, as outlined in~\cite{rebentrost2019quantum}, addresses the problem of gradient descent for a multi-valued input function $f$, aiming to determine the vector at which $f$ is minimized. Both original quantum algorithms exhibit certain speedup relative to the input dimension, compared to best-known classical algorithms, but suffer from expensive resource scaling with respect to other factors. The primary bottleneck in both algorithms arises from the necessity of generating a sequence of desired quantum states through complicated quantum operations and repeated measurements. This inevitably results in a costly process that requires multiple repetitions of the same quantum circuit to obtain the desired states as the input for subsequent computation.
We show that by treating the pre-measured states as a block-encoded operator, simple linear algebraic techniques from the QSVT framework—such as multiplication and linear combination—allow us to directly manipulate these states without the need for the intermediate repeated measurements. This approach eliminates the costly steps imposed in the above two original proposals.
We remark that while the work in~\cite{chakraborty2018power} relies on a complex algorithmic procedure and delicate analysis to achieve an improved quantum algorithm, our findings demonstrate that even elementary operations utilizing the simplest techniques derived from the original QSVT framework~\cite{gilyen2019quantum} can suffice to construct an elegant yet powerful quantum algorithm. These operations enable the removal of the exponential scaling factor present in the original versions of the algorithms~\cite{nghiem2022quantum, rebentrost2019quantum}, as we will elaborate in subsequent sections. }

\revise{The remainder of this work is structured as follows. Section~\ref{sec: overview} provides an overview of the main computational objectives, along with a discussion of two prior proposals~\cite{nghiem2022quantum, rebentrost2019quantum} that address their respective problems. We explicitly highlight the key challenges associated with these algorithms, which serve as the central motivation for our approach. This section also includes a summary of our proposed method, outlining its key ideas and underlying principles.
We then present our improved quantum algorithm for eigenvalue estimation in Section~\ref{sec: eigenvalue}. A concrete comparison is provided to demonstrate the improvements introduced by our approach, which stem from the unitary block encoding framework.
In Section~\ref{sec: gradientdescent}, we address the quantum gradient descent problem~\cite{rebentrost2019quantum}. Following the same structure, we first review the original problem and the solution proposed in~\cite{rebentrost2019quantum}. We then introduce our improved version, which is explored from two different perspectives, followed by a comparison with the original approach.
Finally, in Section~\ref{sec:discussconclude}, we conclude our work with further remarks and discuss potential extensions of our methods to solve a broader range of problems.
}.

\section{Overview of Prior Works and Technical Summary  }
\label{sec: overview}
\revise{In this section, we first provide an overview of the two previous works~\cite{nghiem2022quantum} and~\cite{rebentrost2019quantum}, including the main objective of intere,st as well as a summary of key steps introduced there. From the description, we shall see the main obstacles possessed by these original constructions. Then, we point out the key improvements in this work by providing a technical summary of how our algorithms execute, from which we will see how our proposal circumvents the limitation in~\cite{nghiem2022quantum} and~\cite{rebentrost2019quantum}. More technical details are provided in subsequent discussions, see, e.g., Section~\ref{sec: eigenvalue} and Section~\ref{sec: gradientdescent}. We remark that revelant definitions, as well as elementary tools employed in this work are summarized in Section~\ref{sec: preliminaries}. }

\subsection{Power Method For Finding Largest Eigenvalue}
\label{sec: overviewquantumpowermethod}
\revise{The formal statement of the main objective here is as follows.}\\

\revise{
\noindent\textbf{Problem:} \textit{Given an oracle access to entries of a Hermitian, $s$-sparse matrix $A$ of dimension $n\times n$, whose eigenvalues' norm has known bounds, we estimate its largest eigenvalue (in magnitude) $\lambda_{\max}$ up to an additive error $\delta$} }. \\

\revise{Classically, a popular method for the above problem is the power method~\cite{golub2013matrix1}. The algorithm begins with some randomized vector $x_0$ (the norm does not matter), then applies matrix $A$ to $x_0$ $k$ times, resulting in $x_k = A^k x_0$. The approximation to the largest eigenvalue is given by $\lambda_k = \frac{1}{||x_k||^2} x_k^T A x_k $. Roughly speaking, according to the analysis provided in~\cite{golub2013matrix1} (see Appendix \ref{sec: powermethoditeration}), for $k =\mathcal{O}\Big( \frac{1}{\Delta} \log \frac{1}{\delta} \Big)$ (where $\Delta$ is the difference between two largest eigenvalues in magnitude), then the quantity $\frac{1}{||x_k||^2} x_k^T A x_k $ is $\delta$ close to $\lambda_{\max}$, i.e., $|\lambda_k - \lambda_{\max} |\leq \delta $.  }
\begin{figure}[H]
    \begin{center}
    \begin{tikzpicture}
        % Axes
        \draw[->] (-1,0) -- (5,0) node[right] {$x$};
        \draw[->] (0,-1) -- (0,5) node[above] {$y$};

        % Eigenvector direction
        \draw[thick,red,->] (0,0) -- (4.2,1.8) node[above right] {$v_1$};

        % Initial vector x_0
        \draw[blue,->] (0,0) -- (2,4) node[above] {$x_0$};

        % Iterations
        \draw[blue,->] (0,0) -- (3, 3) node[above] {$x_1$};
        \draw[blue,->] (0,0) -- (3.5, 2.5) node[above] {$x_2$};
        \draw[blue,->] (0,0) -- (3.8, 2.2 ) node[above right] {$x_3$};
        
        % Dotted lines showing alignment
        %\draw[dashed] (1,2) -- (2.5,2);
        %\draw[dashed] (1.5,2.2) -- (2.5,2);
        %\draw[dashed] (2,2.1) -- (2.5,2);
        %\draw[dashed] (2.3,2.05) -- (2.5,2);
    \end{tikzpicture}
\end{center}
    \caption{Illustration of the Power Method. As the matrix $A$ is iteratively applied to an initial vector $x_0$, the resulting vectors $x_1,x_2,x_3, ...$ are gradually aligning with $v_1$, which is the eigenvector corresponding to the largest eigenvalue (in magnitude) of $A$. }
    \label{fig: powermethod}
\end{figure}
\revise{
A further and somewhat standard assumption (also appeared in the linear system context~\cite{harrow2009quantum}) on matrix $A$ is that its eigenvalues' norms are within a fixed range, e.g., $(1/\kappa, 1)$, which is always achievable by a trivial scaling. In~\cite{nghiem2022quantum}, the authors propose a quantum algorithm for the above problem based upon the classical power method, which mainly relies on an iterative multiplication of matrix $A$ to some initial random vector, denoted as $\ket{x_0}$. More specifically, they used the technique introduced in~\cite{wiebe2012quantum}, which is a modified version of the quantum linear solver~\cite{harrow2009quantum}, i.e., matrix multiplication, to make use of the oracle access to entries of $A$ so as to perform the following unitary:
\begin{align}
    U_{A^k} \ket{0}\ket{x_0} &= \ket{0} A^k \ket{x_0} + \ket{1} \ket{\rm Garbage}, \\
                            &=  \ket{0} x_k + \ket{1}\ket{\rm Garbage},
    \label{xk}
\end{align}
where $\ket{\rm Garbage}$ refers to some irrelevant state and $x_k \equiv A^k \ket{x_0}$. Then, measurement is made on the ancilla with post-selection being $\ket{0}$, so as to obtain the state $\ket{x_k} \equiv x_k/||x_k||$, where $||.||$ refers to the usual $l_2$ Euclidean norm. The central quantity of interest to us is:
\begin{align}
    \lambda = \frac{1}{||x_k||^2} x_k^T A x_k  = \bra{x_k}A\ket{x_k}, 
\end{align}
which gives an approximation of the largest eigenvalue. We note that the above quantity is also equivalent to $\lambda = \Tr\Big( A\ket{x_k}\bra{x_k}\Big) $. Hence, given that we already obtain $\ket{x_k}$ from the above measurement, another round of matrix multiplication gives
\begin{align}
    U_A \ket{0}\ket{x_k} = \ket{0} A \ket{x_k} + \ket{1} \ket{\rm Garbage}.
    \label{4}
\end{align}
By obtaining another $\ket{x_k}$, which again is conditioned on the ancilla being measured to $|0\rangle$, and add another ancilla $\ket{0}$, as mentioned in~\cite{nghiem2022quantum}, the overlap $\bra{0} \bra{x_k} U_A \ket{0}\ket{x_k} = \bra{x_k} A \ket{x_k}$ can be estimated, for example, via the Hadamard test, and thereby the estimation $\Tilde{\lambda}$ of the largest eigenvalue can be achieved.
We observe that from the above description, when the measurement is made on the ancilla (of Eqn.~\ref{4}), the probability of measuring $\ket{0}$ is $||x_k||^2 = || A^k \ket{x_0}||^2$. According to the analysis provided in~\cite{nghiem2022quantum}, such a quantity is lower bounded by $\mathcal{O}(\kappa^k)$, and hence the number of measurements required would be exponentially large. Thus, even though the aforementioned algorithm is simple in essence, the running time has an inevitably scaling factor of $\mathcal{O}( \kappa^k )$, where $\kappa$ is the conditional number of given matrix and $k$ is the number of iteration steps. The bottleneck of this approach is a direct consequence of the requirement to obtain the desired state by measurements, which incurs a substantially large probability of failure at each round. \\
\indent To overcome such a costly requirement, we hinge on the observation that, instead of dealing with the post-selected state, we manipulate directly on the pre-measured state, and extract the desired outcome from classical post-processing after the final measurements on the ancilla. We have collected all relevant definitions and tools in Appendix.\ref{sec: preliminaries}. Below, we outline the essential idea of our proposal.
\begin{method}[Improved Algorithm for Eigenvalue Finding]
\label{method: algorithm1}
\end{method}
Input: Oracle access to entries of $s$-sparse, Hermitian matrix $A$ of size $n \times n$. 
\begin{itemize}
    \item Oracle access to entries of $A$ allows us to block encode $A/s$ (see definition \ref{def: blockencode} and Lemma \ref{lemma: As})
    \item From the block encoding of $A/s$, use Lemma \ref{lemma: product} to construct the block encoding of $(A/s)^k$. 
    \item Denote $U$ as the unitary to block encode $(A/s)^k$, $x_k \equiv A^k \ket{x_0}$, $\ket{x_k} = x_k/|x_k|$,  and $p_k = |x_k|^2/s^k$. According to Definition \ref{def: blockencode} and Eqn. \ref{eqn: action}:
    \begin{align}
    U \ket{\bf 0}\ket{x_0} &= \ket{\bf 0} \frac{A^k}{s^k}\ket{x_0}  + \sum_{ \textbf{j} \neq \textbf{0}} \ket{\bf j} \ket{\rm Garbage}_j \\
    &= \sqrt{p_k} \ket{\bf 0}\ket{x_k} + \sum_{ \textbf{j} \neq \textbf{0}} \ket{\bf j} \ket{\rm Garbage}_j.
\end{align}
    \item Appending another ancilla qubit initialized in $\ket{1}$, we have the state $ \sqrt{p_k} \ket{1}\ket{\bf 0}\ket{x_k} + \ket{1}\ket{\rm Garbage} $. Using $\ket{\bf 0}$ as controlled qubits and rotate the ancilla to obtain the following state from the above state:
    \begin{align}
        \sqrt{p_k} \ket{1}\ket{\bf 0}\ket{x_k} + \ket{1} \sum_{ \textbf{j} \neq \textbf{0}} \ket{\bf j} \ket{\rm Garbage}_j  \longrightarrow \sqrt{p_k} \ket{0} \ket{\bf 0}\ket{x_k} + \ket{1}\sum_{ \textbf{j} \neq \textbf{0}} \ket{\bf j} \ket{\rm Garbage}_j\equiv \ket{\Phi}.
    \end{align}
    \item Tracing out the register holding $\ket{\bf 0}, \ket{\bf j}$, we obtain the density state: 
    \begin{align}
         p_k \ket{0}\bra{0}\otimes \ket{x_k}\bra{x_k} + \ket{1}\bra{1}\otimes \sum_{\textbf{j} \neq \textbf{0}} \ket{\rm Garbage}_j\bra{\rm Garbage}_j.
    \end{align}
    The above density matrix can be block-encoded by virtue of Lemma~\ref{lemma: improveddme}. We remark  that the above operator is again a block encoding of $ p_k \ket{x_k}\bra{x_k}$. 
    \item Prepare a block encoding of $  \ket{\phi}\bra{\phi}$, where $\ket{\phi}$ is some randomly known state, which is possible by Lemma~\ref{lemma: improveddme}. Then use Lemma~\ref{lemma: product} to construct the block encoding of 
    \begin{align}
        p_k \ket{x_k}\bra{x_k} \cdot \ket{\phi}\bra{\phi}  =  p_k \braket{x_k,\phi} \ket{x_k}\bra{\phi}.
    \end{align}
    \item Use Lemma \ref{lemma: theorem56} to transform the above block-encoded operator into 
    \begin{align}
        \exp\Big( -\beta\big( 1- p_k\braket{x_k,\Phi} \big)  \Big)  \ket{x_k}\bra{\phi}.  
        \end{align}
    \item Prepare the state $\ket{\bf 0} \ket{\phi}$. Use the block encoding of the above operator and apply to the state $\ket{\bf 0} \ket{\phi}$ and use the property of Def.~\ref{def: blockencode} as well as Eqn.~\ref{eqn: action}, we obtain:
    \begin{align}
      \ket{\bf 0} \exp\Big( -\beta\big( 1- p_k\braket{x_k,\phi} \big)  \Big) \ket{x_k} + \sum_{ \textbf{j} \neq \textbf{0}} \ket{\bf j} \ket{\rm Garbage}_j.
    \end{align}
    \item Use amplitude estimation technique to estimate the value $\exp\Big( -\beta\big( 1- p_k\braket{x_k,\phi} \big)  \Big) $.
    \item Append another ancilla initialized in $\ket{1}$, we obtain the state: 
    \begin{align}
        \ket{1} \ket{\bf 0} \exp\Big( -\beta\big( 1- p_k\braket{x_k,\phi} \big)  \Big) \ket{x_k} + \ket{1}\sum_{ \textbf{j} \neq \textbf{0}} \ket{\bf j} \ket{\rm Garbage}_j.
    \end{align}
    \item Use $\ket{\bf 0}$ as the controlled qubits and apply $X$ gate on the first qubit,  we obtain:
    \begin{align}
         \ket{0} \ket{\bf 0} \exp\Big( -\beta\big( 1- p_k\braket{x_k,\phi} \big)  \Big) \ket{x_k} + \ket{1} \sum_{ \textbf{j} \neq \textbf{0}} \ket{\bf j} \ket{\rm Garbage}_j.
    \end{align}
    \item Tracing out the register that holds $\ket{\bf 0},\ket{\bf j}$, we obtain the density matrix:
    \begin{align}
       \Phi = \ket{0}\bra{0} \otimes \exp\Big( -2\cdot\beta\big( 1- p_k\braket{x_k,\phi} \big)  \Big)\ket{x_k}\bra{x_k} + \ket{1}\bra{1}\otimes  \sum_{ \textbf{j} \neq \textbf{0}}  \ket{\rm Garbage}_j \bra{\rm Garbage}_j.
    \end{align}
    \item Prepare some known 1-qubit gate $M_1$, and use Lemma~\ref{lemma: tensorproduct} to construct the block encoding of $M \otimes A/s$. Leverage the result of~\cite{rall2020quantum} and estimate:
    \begin{align}
        &\Tr\Big( \big( M_1\otimes \frac{A}{s}\big) \cdot \Phi \Big) \\
        &= \exp\Big( -2\cdot\beta\big( 1- p_k\braket{x_k,\phi} \big)  \Big) \Tr\big( M_1 \ket{0}\bra{0}\big)\frac{1}{s} \Tr\Big( A\ket{x_k}\bra{x_k} \Big) + \Tr\big( M_1\ket{1}\bra{1}\big)  \Tr\Big( \frac{A}{s} \sum_{ \textbf{j} \neq \textbf{0}}  \ket{\rm Garbage}_j \bra{\rm Garbage}_j  \Big).
    \end{align}
    \item By choosing another 1-qubit gate $M_2$, and repeat the same procedure, we estimate:
     \begin{align}
        &\Tr\Big( \big( M_2\otimes \frac{A}{s}\big) \cdot \Phi \Big) \\
        &= \exp\Big( -2\cdot\beta\big( 1- p_k\braket{x_k,\phi} \big)  \Big) \Tr\big( M_2 \ket{0}\bra{0}\big)\frac{1}{s} \Tr\Big( A\ket{x_k}\bra{x_k} \Big) + \Tr\big( M_2\ket{1}\bra{1}\big)  \Tr\Big( \frac{A}{s} \sum_{ \textbf{j} \neq \textbf{0}}  \ket{\rm Garbage}_j \bra{\rm Garbage}_j  \Big).
    \end{align}
    \item Defining
    \begin{align}
        \lambda =\Tr\Big( A\ket{x_k}\bra{x_k} \Big), \gamma =  \Tr\Big( \frac{A}{s} \sum_{ \textbf{j} \neq \textbf{0}}  \ket{\rm Garbage}_j \bra{\rm Garbage}_j  \Big),\\
        a_{11} = \Tr\big( M_1 \ket{0}\bra{0}\big), a_{12} = \Tr\big( M_1 \ket{1}\bra{1}\big),\\
        a_{21} = \Tr\big( M_2 \ket{0}\bra{0}\big), a_{22} = \Tr\big( M_2 \ket{1}\bra{1}\big),
    \end{align}
    then we obtain a linear system:
    \begin{align}
        \begin{cases}
          \exp\Big( -2\cdot \beta\big( 1- p_k\braket{x_k,\phi} \big)  \Big)  a_{11} \lambda + a_{12} \gamma = \Tr\Big( \big( M_1\otimes \frac{A}{s}\big) \cdot \Phi \Big),\\
           \exp\Big( -2 \cdot \beta\big( 1- p_k\braket{x_k,\phi} \big)  \Big) a_{21} \lambda + a_{22} \gamma = \Tr\Big( \big( M_2\otimes \frac{A}{s}\big) \cdot \Phi \Big).
        \end{cases}
    \end{align}
    The value of $\lambda$ can be revealed by solving the above linear system, as the value of $\exp\Big( -\beta\big( 1- p_k\braket{x_k,\phi} \big)  \Big)$ has been found previously. 
\end{itemize}
Output: $\lambda = \bra{x_k} A \ket{x_k}$.
}

\smallskip
\revise{From the summary above of the algorithm, we have seen that there is no measurement needed in the intermediate step to obtain the desired state $\ket{x_k}$. Thus, our proposed method here removes the exponential scaling mentioned earlier and, in turn, provides an efficient algorithm with significantly better scaling, which will be shown in greater details in Section~\ref{sec: eigenvalue}, consequently providing a major improvement over the method introduced in~\cite{nghiem2022quantum}. } 

\subsection{Gradient Descent for Finding Extrema}
\label{sec: overviewquantumgradientdescent}
\revise{In the second problem, we explore how the unitary block encoding method can be used to improve the quantum gradient descent method proposed in~\cite{rebentrost2019quantum}, which is formally stated as follows.} \\

\noindent\textbf{Problem:} \textit{Find $x$ that minimizes a function $f: \mathbb{R}^n \rightarrow \mathbb{R}$,  where $f$ is some homogeneous polynomial of even degree. }\\
\begin{figure}[H]
    \begin{center}
    \begin{tikzpicture}
        % Define 3D plot
        \begin{axis}[
            view={45}{45},
            xlabel={$x$}, ylabel={$y$}, zlabel={$f(x,y)$},
            colormap/viridis,
            grid=major,
            domain=-2:2,
            y domain=-2:2,
            samples=20,
        ]
            % Surface plot of a quadratic function
            \addplot3[surf,opacity=0.7] {x^2 + y^2};

            % Gradient descent path
            \addplot3[color=red, thick, mark=*] coordinates {
                (1.8,1.8,6.48)
                (1.4,1.4,3.92)
                (1.0,1.0,2.0)
                (0.6,0.6,0.72)
                (0.3,0.3,0.18)
                (0.1,0.1,0.02)
                (0,0,0)
            };
        \end{axis}
    \end{tikzpicture}
\end{center}
    \caption{Illustration of gradient descent method for a two-variables function. Beginning at some initial guess point $\xbf_0$, in an iterative manner, the algorithm gradually moves along the direction of the steepest gradient (see the red dots and line showing the trajectory). The method typically stops if further iteration no longer ``moves'' the point, indicating that the gradient is approaching zero, which means that it has reached the extrema.  }
    \label{fig: illugradient}
\end{figure}
\revise{
A very popular method to solve the above problem is the gradient descent method. According to~\cite{rebentrost2019quantum}, we begin with some random guess $x_0$, and then iteratively perform the following subtraction: 
\begin{align}
    x_{t+1} = x_t  - \eta \bigtriangledown f(x_t),
\end{align}
where $t$ refers to the iteration step and $\eta$ is regarded as the time interval of integration. In~\cite{rebentrost2019quantum}, the authors imposed the spherical constraint, i.e., $x^T x = 1$,  where $x \in \mathbb{R}^n$ refers to the input of $f$. The spherical constraint simply implies that the temporal solution $x_{t+1}$ is normalized after each iteration before performing the next step. It means that the above step should be rewritten as:
\begin{align}
    x_{t+1} = \ket{x_t} - \eta \bigtriangledown f( \ket{x_t}),
\end{align}
and, subsequently, one uses the normalized state $|x_{t+1}\rangle$ as a starting point and repeats the process.}

\revise{The challenge in this method is the evaluation of the gradient of the function $f$. Let us briefly mention some definitions from the original work~\cite{rebentrost2019quantum} that we will need.  Recall that in~\cite{rebentrost2019quantum}, the authors considered a real-valued function $f$ that is a polynomial of degree $2p$:
\begin{align}
    f(x) = \frac{1}{2} \sum_{m_1,m_2,...,m_p=1}^n A_{m_1, m_2,...,m_{2p}} x_{m_1}... x_{ m_{2p} }. 
\end{align}
Particularly, it can be written in an algebraic form: 
\begin{align}
\label{eqn: algebraicform}
    f(x) = \frac{1}{2} \bra{x} \otimes \cdots \otimes \bra{x} A \ket{x} \otimes \cdots \otimes \ket{x},
\end{align}
where $A$ is a real matrix of dimension $n^p \times n^p$ with a bounded norm, and with sparsity $s$. Furthermore, $A$ can be formally decomposed as: 
\begin{align}
    A = \sum_{\alpha =1}^K A^{\alpha}_1 \otimes A^{\alpha}_2 \cdots \otimes A^{\alpha}_p,
\end{align}
where $K$ is some number counting the terms in the decomposition and each $A^{\alpha}_i$ is a matrix of dimension $n \times n$ (which can be a distinct matrix for each $i$). Given the above tensor formulation, the gradient of $f$ can be rewritten as:
\begin{align}
    \bigtriangledown f(x) = D(\ket{x}) \ket{x},
\end{align}
where 
\begin{align}
    D(\ket{x}) = \sum_{\alpha=1}^K\sum_{m=1}^p \Big( \prod_{n=1, n\neq m}^p  \bra{x} A_n^{\alpha} \ket{x} \Big) A_m^{\alpha}.
    \label{eqn: Dop} 
\end{align}
It is useful to note that the above gradient operator $D$ is equivalent to: 
\begin{align}
    D = tr_{1,2,...,p-1} ( \ket{x}\bra{x} ^ {\otimes (p-1)} \otimes \mathbb{I}) M_D,
    \label{eqn: D}
\end{align}
where
\begin{align}
\label{eq:MD}
    M_D = \sum_{\alpha=1}^K\sum_{m=1}^p \Big( \bigotimes_{n=1,n\neq  m}^p A_{n}^{\alpha} \Big) \bigotimes A_m^{\alpha}. 
\end{align}
As further shown in~\cite{rebentrost2019quantum}, the above $M_D$ can be simplified as 
\begin{align}
    M_D = \sum_{j=1}^p M_j = \sum_{j=1}^p Q_j A Q_j,
    \label{eqn: md}
\end{align}
where $Q_j$ is the swap operation between the $j$-th register and the final register. Therefore, the oracle access to $A$ can be used to construct $M_j$ for all $j$.} 

\revise{We refer the readers to the original work~\cite{rebentrost2019quantum} for full details of the execution of the quantum gradient descent, as well as the cost of the algorithm. Here, we simply point out that the gradient operator $D$ (see Eqn.~\ref{eqn: D}) literally consumes multiple copies of $\ket{x_t}$ at each of the $t$-th iteration step. As $M_D$ is not unitary, it is not straightforward to execute $D$. What the authors in~\cite{rebentrost2019quantum} did was that they employed the Hamiltonian simulation method to simulate $\exp(-i M_D \delta t )$ for varying time steps $\delta t$, adapting the quantum state exponentiation procedure outlined in~\cite{lloyd2014quantum}, e.g., as in the following approximation:
\begin{align}
  \Tr_{1,2,...,p-1} \Big( \exp(-i M_D \delta t) \ket{x}\bra{x}^{\otimes p} \exp(i M_D \delta t) \Big)  \approx \exp(-iD \delta t) \ket{x}\bra{x} \exp(i D \delta t).
  \label{eqn: trick}
\end{align}
From the simulation of $\exp(-i D \delta t)$, a similar method as in~\cite{wiebe2014quantum} is employed to finally obtain the multiplication $D$ to some state $\ket{x}$. While we have left out a lot of technical details in the above description, there are two main obstacles. First, the simulation step $\exp(-i M_D \delta t)$, plus the above partial trace step and the final subroutine for applying $D$, uses Hamiltonian simulation and density matrix exponentiation technique~\cite{lloyd2013quantum, berry2007efficient,berry2012black,berry2014high}, which incurs a high scaling on error tolerance. The second is, as we have seen above, at each iteration step, there are $p$ copies of state $\ket{x}$ required, (where $\ket{x}$ is obtained from the previous procedure), which incurs a significant amount of measurement. This issue possesses similarity to the difficulty of obtaining the desired state $\ket{x_k}$, e.g., see Equation~\ref{xk}. Thus, we expect that the inspiration from our proposal for the improved quantum power method can be adopted.
}

\revise{Indeed, the solution to the above difficulty turns out to be somewhat similar to what we had for the largest eigenvalue finding, as instead of focusing on measurement/post-selection, we aim to treat the vector (or quantum state) of interest in the density matrix formalism, and leverage the technique of block-encoding to carry out the arithmetic operations.
In this part, we introduce \textit{two} alternative new approaches to the gradient descent. \smallskip \\
\noindent
\textbf{First Version:} 
It is an indirect improvement to the original work in~\cite{rebentrost2019quantum}, and we make an adjustment to the input and output of the quantum Newton method. Roughly speaking, instead of working with vectors $\Vec{x}$ (pure states in the context of~\cite{rebentrost2019quantum}), we work with the density matrix form $\Vec{x}\Vec{x}^T$. By recasting this way, our algorithm, once beginning with an initial guessed operator $x_0 x_0^T$, allows a simple iterative procedure that leads to the final operator $x_T x_T^T$ corresponding to the final solution vector of the Newton method for a total of $T$ iterations (note $T$ in the subscript of $x$). This approach admits a highly efficient scaling in the error tolerance $\epsilon$, which is polylogarithmic. On the contrary, the original work~\cite{rebentrost2019quantum} has time scaling dependence being polynomial in $1/\epsilon$. Below, we also point out how the same idea (of this version) can be applied to the problem of finding the inverse of a matrix. To be more concise, the central ideas of our method proceed as follows:
\begin{method}[Alternative Quantum Gradient Descent Algorithm]
\label{method: algorithm2}
\end{method}
Input: Oracle access to entries of $A$ that defines the homogeneous polynomial of even degree (Eqn.~\ref{eqn: algebraicform})
\begin{itemize}
    \item Treat a vector $x$ as operator $x x^\dagger$ block-encoded via some unitary. 
    \item Use Lemma \ref{lemma: tensorproduct} to construct the block encoding of $ (x x^\dagger)^{\otimes p-1} \otimes \mathbb{I}$ (note that as $\mathbb{I}$ is trivial, its unitary block encoding is also trivial).
    \item Use the oracle access to all $\{ M_j \}_{j=1}^p$ (see Eqn.~\ref{eq:MD}) to construct the block encoding of all $\{ M_j /s \}_{j=1}^p$ (via Lemma~\ref{lemma: As}). 
    \item From the above, use Lemma \ref{lemma: sumencoding} to construct $\sum_{j=1}^p M_j/(ps) = M_D/(ps)$. 
    \item Observe the following property: 
    \begin{align}
    (x x^T)^{\otimes (p-1)} \otimes \mathbb{I} \,\,\cdot\frac{M_D}{ps}\,\cdot (x x^T)^{\otimes (p-1)} \otimes \mathbb{I} 
    =  (x x^T)^{\otimes p-1} \otimes \frac{D(x)}{ps}.
    \label{21}
\end{align}
    \item From the above block encoding, use Appendix~\ref{sec: proof11} to obtain the block encoding of $D(x)/ps$. 
    \item Use Lemma \ref{lemma: sumencoding} and also Lemma~\ref{lemma: scale} to carry out the transformation of block-encoded operators:
    \begin{align}
        x x^\dagger \longrightarrow \Big( \frac{x}{ps} - \eta \frac{D(x)}{ps}   \Big )\Big( \frac{x^\dagger}{ps} - \eta \frac{D(x)^\dagger}{ps}\Big).
    \end{align}
    We remark that the above step is nothing but the gradient descent step with hyperparameter $\eta$. 
    \item Repeat all the above  steps with a total of $T$ iterations. 
\end{itemize}
Output: A block encoding of the operator $\xbf_T\xbf_T^\dagger$. \smallskip\\}

\revise{
\noindent
\textbf{Second version:} 
Here, we make a direct improvement on the original algorithm proposed in Ref.~\cite{rebentrost2019quantum}, as we keep the quantum state itself as the main working object. It means that at the $t$-th iteration step, the input is (multiple copies of) $\ket{x_t}$ and we need to produce the output $\ket{x_{t+1}}$. Our proposal has the following key ideas.
\begin{method}[Improved Quantum Gradient Descent Algorithm]
\label{method: algorithm3}
\end{method}
Input: Oracle access to entries of $A$ that defines the homogeneous polynomial of even degree (Eqn.~\ref{eqn: algebraicform})
    \begin{itemize}
        \item Take multiple copies of $\ket{x_t}$ and construct the exponentiation $\exp(-i \ket{x_t}\bra{x_t}t ) \equiv \exp(-i\rho_t )$ using technique introduced in~\cite{lloyd2013quantum}.
        \item Use a tool from~\cite{gilyen2019quantum} (Corollary 71, see Lemma~\ref{lemma: logU} below) to construct the block encoding of $\pi\rho_t /4 $ (the factor $\pi/4$ appears due to technical step). 
        \item Use Lemma~\ref{lemma: tensorproduct} to construct the block encoding of $(\frac{\pi}{4}\rho_t)^{\otimes p-1} $ and of $(\frac{\pi}{4}\rho_t)^{\otimes p}$. 
        \item From the oracle access to entries of $A$, construct the block encoding of $M_D/ps$. 
        \item Use Lemma~\ref{lemma: product} to construct the block encoding of:
        \begin{align}
            \Big( \big(\frac{\pi}{4}\rho_t\big)^{\otimes p-1} \otimes \Ibb \Big) \frac{M_D}{ps} \Big( \big(\frac{\pi}{4}\rho_t\big)^{\otimes p}  \Big) = \big(\frac{\pi}{4}\big)^{2p-1} (\rho_t)^{\otimes p-1} \otimes \frac{D(\ket{x_t}) }{ps} \rho_t,
        \end{align}
        where we have used a property similar to Eqn.~\ref{21} which will be proved in Appendix~\ref{sec: proof11}.
        \item Use Lemma~\ref{lemma: scale} to transform the above into the block encoding of 
        $$ \big(\frac{\pi}{4}\big)^{2p-1} \rho_t^{\otimes p-1} \otimes \eta \frac{D(\ket{x_t}) }{ps} \rho_t. $$
        \item Take the block encoding of $(\frac{\pi}{4}\rho_t)^{\otimes p}$ and use Lemma~\ref{lemma: scale} to construct the block encoding of $ \big(\frac{\pi}{4}\big)^{2p-1}(\rho_t)^{\otimes p}/ps$. Then we can use Lemma~\ref{lemma: sumencoding} to construct the block encoding of
        \begin{align}
            \big(\frac{\pi}{4}\big)^{2p-1}\frac{\rho_t^{\otimes p}}{ps}  - \big(\frac{\pi}{4}\big)^{2p-1} \rho_t^{\otimes p-1} \otimes \frac{\eta D(\ket{x_t}) }{ps} \rho_t = \big(\frac{\pi}{4}\big)^{2p-1} \frac{\rho_t^{p-1}}{ps}  \otimes \Big(  \rho_t - \eta D(\ket{x_t}) \rho_t \Big).
        \end{align}
        \item Denote the above unitary block encoding as $U$. Use the property of block encoding (see Definition~\ref{def: blockencode} and property in Equation~\ref{eqn: action}) and perform:
        \begin{align}
            U \ket{\bf 0}\ket{x_t}^{\otimes p} &= \ket{\bf 0} \ket{x_t}^{\otimes p-1} \big(\frac{\pi}{4}\big)^{2p-1}\frac{1}{ps} ( \ket{x_t} - \eta D(\ket{x_t}) \ket{x_t} ) + \ket{\rm Garbage} \\
            &= \ket{\bf 0} \ket{x_t}^{\otimes p-1} \big(\frac{\pi}{4}\big)^{2p-1}\frac{1}{ps}  x_{t+1} + \ket{\rm Garbage},
        \end{align}
        where we have used the fact that $\rho_t \equiv \ket{x_t}\bra{x_t} $ and that $ \rho_t \ket{x_t} = \ket{x_t}$. Additionally, we have defined the following un-normalized vector
        $$ x_{t+1} =\ket{x_t} - \eta D(\ket{x_t}) \ket{x_t}. $$ 
        \item Perform measurement on the first ancilla, post-selecting on $\ket{\bf 0}$ and observe on the last register, we obtain the state $\ket{x_{t+1}}$. We remark that the success probability in this case, as shown in~\cite{rebentrost2019quantum}, is lower bounded by a constant. 
    \end{itemize}
Output: a quantum state $\ket{\xbf_{t+1}}$ corresponding to $\xbf_{t+1}$--which is an updated solution. \\
For such a single step in the gradient descent method, our algorithm requires a power-of-three fewer number of copies of the given state at step $t$. Furthermore, the time complexity of our approach has a power-of-four advantage compared to that of Ref.~\cite{rebentrost2019quantum} with respect to the error tolerance $\epsilon$. }

\section{Improving Eigenvalues Estimation}
\label{sec: eigenvalue}
In this section we provide a detailed version of Algorithm~\ref{method: algorithm1}, with each step being explicitly presented and explained. 

\revise{The first recipe we need is the following result.} 
\begin{lemma}
\label{lemma: newmatrixapp}
Given oracle access to some $s$-sparse matrix $A$, then the following unitary:
\begin{align}
        U_{A_k} \ket{\bf 0}\ket{x_0} = \ket{\bf 0} \frac{A^k}{s^k} \ket{x_0} + \sum_{j \neq \textbf{0} } \ket{j}\ket{\rm Garbage}_j   
        \label{eqn: newmatrixapp}
\end{align}
can be realized up to additive accuracy $\epsilon$ in time 
$$ \mathcal{O}\left( k \Big(\log(n) + \log^{2.5}\big(\frac{sk}{\epsilon}\big)  \Big)\right), $$
where $\ket{\bf 0}$ refers to an ancillary system. We note that all $\{\ket{\rm Garbage}_j\}$ are not properly normalized.
\end{lemma}
The proof of the above lemma is simple. We use Lemma \ref{lemma: As} to obtain an $\epsilon$-approximated block encoding of $A/s$. Note that due to the error accumulation, we require each block encoding of $A/s$ to have error tolerance $\epsilon/k$. We then use Lemma~\ref{lemma: product} $k$ times to obtain an $\epsilon$-approximated block encoding of $(A/s)^k$. By using the definition of the block encoding (see Def \ref{def: blockencode} and Eqn. \ref{eqn: action}), we have arrived at equation~(\ref{eqn: newmatrixapp}). 

For convenience, we write the equation (\ref{eqn: newmatrixapp}) as:
\begin{align}
     U_{A_k} \ket{\bf 0}\ket{x_0} &= \ket{\bf 0} \frac{A^k}{s^k} \ket{x_0} +  \revise{ \sum_{ \textbf{j} \neq \textbf{0}} \ket{\bf j} \ket{\rm Garbage}_j }\\  \nonumber  \\
     &= \sqrt{p_k} \ket{\bf 0}\ket{x_k} +  \revise{ \sum_{ \textbf{j} \neq \textbf{0}} \ket{\bf j} \ket{\rm Garbage}_j },
     \label{eqn: newstate}
\end{align}
where $x_k = A^k \ket{x_0}$ and $\ket{x_k}$ is the corresponding normalized quantum state, and $p_k = ||x_k||^2/s^k$. \revise{Now we append an ancilla initialized in $\ket{1}$, which can be simply obtained by apply $X$ gate to the state $\ket{0}$. So we obtain the following state:}
\revise{\begin{align}
    \sqrt{p_k} \ket{1}\ket{\bf 0}\ket{x_k} +  \ket{1}\sum_{ \textbf{j} \neq \textbf{0}} \ket{\bf j} \ket{\rm Garbage}_j. 
\end{align}}
\revise{By using $\ket{\bf 0}$ as a controlled qubit to apply $X$ gate on the first ancilla qubit, we obtain the state:}
\revise{\begin{align}
    \sqrt{p_k} \ket{0}\ket{\bf 0}\ket{x_k} +  \ket{1}\sum_{ \textbf{j} \neq \textbf{0}} \ket{\bf j} \ket{\rm Garbage}_j. 
\end{align}}
\revise{If we trace out the second register, that contains $\ket{\bf 0}$, $\ket{\rm Garbage}_j$, from the above state, it results in the following state: }
\revise{\begin{align}
    p_k \ket{0}\bra{0} \otimes \ket{x_k}\bra{x_k} + \ket{1}\bra{1}\otimes \Big( \sum_{ \textbf{j} \neq \textbf{0}}  \ket{\rm Garbage}_{\bf j} \bra{\rm Garbage}_{\bf j} \Big).
\end{align}}
\revise{We remark that the above density matrix is obtained tracing out a quantum state, so it is possibly block-encoded using the following result: 
\begin{lemma}[\cite{gilyen2019quantum}]
\label{lemma: improveddme}
Let $\rho = \Tr_A \ket{\Phi}\bra{\Phi}$, where $\ket{\Phi} \in  \mathbb{H}_A \otimes \mathbb{H}_B$, and thus $\rho$ is a density matrix that acts on states  in $ \mathbb{H}_B$. Given a unitary $U$ that generates $\ket{\Phi}$ from $\ket{\bf 0}_A \otimes \ket{\bf 0}_B$, then there exists a procedure that constructs an exact unitary block encoding of $\rho$ in complexity $\mathcal{O}(T_U + \log(n) )$, where $n$ is the dimension of $\mathbb{H}_B$. 
\end{lemma}
We observe further that the operator
$$ p_k \ket{0}\bra{0} \otimes \ket{x_k}\bra{x_k} + \ket{1}\bra{1}\otimes \Big( \sum_{ \textbf{j} \neq \textbf{0}}  \ket{\rm Garbage}_j \bra{\rm Garbage}_j \Big) $$
is also a block encoding of $ p_k \ket{x_k}\bra{x_k}$ according to Definition~\ref{def: blockencode}. Thus, by virtue of Lemma~\ref{lemma: improveddme}, we naturally obtain the block encoding of $ p_k \ket{x_k}\bra{x_k}$. Now we take a random state $\ket{\phi}$ with a known generating circuit (assumed to have $\mathcal{O}(1)$ depth), and note that $\ket{\phi}\bra{\phi}$ is easily made the block encoding of itself by attaching identity operator. Then we use Lemma~\ref{lemma: product} to construct the block encoding of $p_k \ket{x_k}\bra{x_k} \cdot \ket{\phi}\bra{\phi} = p_k \braket{x_k,\phi} \ket{x_k}\bra{\phi} $. We note that the step in which we use $\ket{\bf 0}$ as controlled qubits and apply $X$ to the ancilla increases the circuit depth by at most $\mathcal{O}(\log n)$, so the circuit complexity for producing an $\epsilon$-approximated block encoding of $p_k \braket{x_k,\phi} \ket{x_k}\bra{\phi}$ is $\mathcal{O}\left( k \Big(\log(n) + \log^{2.5}\big(\frac{sk}{\epsilon}\big)  \Big)\right)$.}

\revise{The block encoding of $p_k\braket{x_k,\phi} \ket{x_k}\bra{\phi}$ has a small coefficient $p_k$, and we aim to solve this problem. To proceed, we mention the following results.
\begin{lemma}\label{lemma: qsvt}[\cite{gilyen2019quantum} Theorem 56]
\label{lemma: theorem56}  (Polynomial Transformation)
Suppose that $U$ is an
$(\alpha, a, \epsilon)$-encoding of a Hermitian matrix $A$. (See Definition 43 of~\cite{gilyen2019quantum} for the definition.)
If $P \in \mathbb{R}[x]$ is a degree-$d$ polynomial satisfying that
\begin{itemize}
\item for all $x \in [-1,1]$: $|P(x)| \leq \frac{1}{2}$,
\end{itemize}
then, there is a quantum circuit $\tilde{U}$, which is an $(1,a+2,4d \sqrt{\frac{\epsilon}{\alpha}})$-encoding of $P(A/\alpha)$ and
consists of $d$ applications of $U$ and $U^\dagger$ gates, a single application of controlled-$U$ and $\mathcal{O}((a+1)d)$
other one- and two-qubit gates.
\end{lemma}}
\revise{
\begin{lemma}
\label{lemma: exponential}[~\cite{gilyen2019quantum}  Corollary 64]  (Polynomial Approximation to Exponential Function)
   Let $\beta \in \mathbb{R}_+$ and $\epsilon \in (0,1/2]$. There exists an efficiently constructible polynomial $P \in \mathbb{R}[x]$ such that 
   $$ \Big|\!\Big| e^{ -\beta ( 1-x ) } - P(x)  \Big|\!\Big|_{x\in[-1,1]} \leq \epsilon. $$
   Moreover, the degree of $P$ is $\mathcal{O}\Big( \sqrt{\max[\beta, \log(\frac{1}{\epsilon})] \log(\frac{1}{\epsilon}}) \Big) $.
\end{lemma} }
\revise{As $p_k\sim 1/s^k$ is exponentially small in $k$, it can potentially cause a problem that measuring the ancillas rarely gives the desired outcome. Recall from Section~\ref{sec: overview} that we eventually produce a small linear system, and solving it yields our desired quantity $\lambda$. The conditional number of the linear systems turns out to be $\sim 1/p_k$, and hence is exponentially large, which leads to a large deviation of errors. Therefore, we need to find a way to amplify $p_k$ to prevent this, and the above two lemmas are the techniques we can use to achieve this. }

\revise{The block encoding of $p_k \braket{x_k,\phi} \ket{x_k}\bra{\phi}$ allows us to leverage Lemma~\ref{lemma: theorem56} and Lemma~\ref{lemma: exponential} to approximately transform the block-encoded operator $p_k \braket{x_k,\phi} \ket{x_k}\bra{\phi}$ into the operator $\exp\Big(- \beta\big( 1-p_k  \braket{x_k,\phi} \big)  \Big) \ket{x_k}\bra{\phi}$ (where we shall choose $\beta = 0.01$ in the above lemma). There are a few subtile issues. In the statement of the above Lemma ~\ref{lemma: theorem56}, the block-encoded matrix is supposed to be Hermitian, but $\ket{x_k}\bra{\phi}$ is not. However, as 
indicated in Theorem 17 of \cite{gilyen2019quantum}, the essentially same procedure (plus 
complexity) still works for a non-Hermitian matrix $A$, but it transforms the singular 
values of $A$ instead of eigenvalues. Next, the transformation is \textit{approximated} because we 
actually use Lemma~\ref{lemma: theorem56} to transform $p_k \braket{x_k,\phi}$ into $P(p_k \braket{x_k,\phi})$, where $P(x)$ is the polynomial approximation of $\exp\big(-\beta (1-x) \big)$, given by the Lemma \ref{lemma: exponential}. One may wonder why we use such an exponential function. We will come back to this point later, as it is crucially related to the stability of the linear system that we will need to solve in order to obtain the desired quantity $\lambda  = \bra{x_k}A\ket{x_k}$. We note that the circuit generating $\ket{\phi}$ has a constant depth (independent of $n$), the quantum circuit for producing an $\epsilon$-approximated block encoding of $\exp\Big(- \beta \big(
1-p_k \braket{x_k,\phi} \big)  \Big) \ket{x_k}\bra{\phi}$ has complexity 
$\mathcal{O}\left(  \Big(\log \frac{1}{\epsilon} \Big) k \Big(\log n + \log^{2.5} \frac{sk}{\epsilon}\Big) \right)$. 
}

\revise{Given that we have obtained the unitary block encoding of $\exp\Big(- \beta \big( 1-p_k \braket{x_k,\phi}   \Big) \ket{x_k}\bra{\phi}$, we apply it to the state $\ket{\bf 0}\ket{\phi}$ (which can be easily prepared), and according to Definition \ref{def: blockencode} and Eqn. \ref{eqn: action}, we thus obtain the state:
\begin{align}
    & \ket{\bf 0}\exp\Big(- \beta \big( 1-p_k \braket{x_k,\phi}\big)   \Big) \ket{x_k}\bra{\phi} \cdot \ket{\phi} + \sum_{\textbf{j} \neq \textbf{0}} \ket{\bf j}\ket{\rm Garbage}_{\bf j} \\
    &= \exp\Big(- \beta \big( 1-p_k \braket{x_k,\phi} \big)  \Big)\braket{x_k,\phi}\ket{\bf 0} \ket{x_k} + \sum_{\textbf{j} \neq \textbf{0}} \ket{\bf j}\ket{\rm Garbage}_{\bf j}.
\end{align}
From the above state, we can first apply amplitude estimation techniques~\cite{rall2023amplitude, aaronson2020quantum, rall2021faster, manzano2023real, grinko2021iterative} to estimate the value of $ \exp\Big(- \beta \big( 1-p_k \braket{x_k,\phi} \big)  \Big)\braket{x_k,\phi}$, which invokes $\mathcal{O}(1/\epsilon)$ uses of the unitary that generates the above state. We append an ancilla qubit initialized in $\ket{1}$, 
\begin{align}
    \exp\Big(- \beta \big( 1-p_k \braket{x_k,\phi} \big)  \Big)\braket{x_k,\phi} \ket{1}\ket{\bf 0} \ket{x_k} + \ket{1} \sum_{\textbf{j} \neq \textbf{0}} \ket{\bf j}\ket{\rm Garbage}_{\bf j},
\end{align}
and then we use the second register as the controlled qubits and apply $X$ on the ancilla (first register) conditioned on it being $\ket{\bf 0}$. We thus obtain:
\begin{align}
    \exp\Big(- \beta \big( 1-p_k \braket{x_k,\phi} \big)  \Big)\braket{x_k,\phi} \ket{0}\ket{\bf 0} \ket{x_k} + \sum_{\textbf{j} \neq \textbf{0}} \ket{1} \ket{\bf j}\ket{\rm Garbage}_{\bf j}.
\end{align}
Tracing out the second register (which holds $\ket{\bf 0}, \ket{\bf j}$), we obtain the density matrix:
\begin{align}
  \rho_k \equiv  \exp\Big(- \beta \big( 1-p_k \braket{x_k,\phi} \big)  \Big)^2 \ket{0}\bra{0}\otimes \ket{x_k}\bra{x_k} + \ket{1}\bra{1}\otimes \sum_{\textbf{j} \neq \textbf{0}}  \ket{\rm Garbage}_{\bf j}  \bra{\rm Garbage}_{\bf j}.
\end{align}
We note that using multi-controlled gates only incurs a modest circuit of depth $\mathcal{O}(\log n)$. The amplitude estimation step incurs a quantum circuit depth
$$\mathcal{O}\left(  \frac{k}{\epsilon} \Big(\log\frac{1}{\epsilon}  \Big) \Big(\log(n) + \log^{2.5}\big(\frac{sk}{\epsilon}\big)  \Big)\right).$$
}

\revise{Next, we need to design a way to read out the desired results to estimate the largest eigenvalue. For this, we employ the following result.}
\begin{lemma}[\cite{rall2020quantum}]
\label{lemma: rall}
    Given the block encoding $U$ of some matrix $A$ (whose norm less than 1) and a unitary $U_\rho$ that satisfies 
    $$  \rho =  \Tr_{E} (U_\rho \ket{\bf 0} \ket{0}_E \bra{\bf 0}\bra{0}_E U_\rho^\dagger), $$ 
    \revise{where the subscript $E$ denote the subsystem that is being traced out}. Then the quantity $ \Tr(A \rho)$ can be estimated up to an additive error $\epsilon$ using a circuit of size 
    $$ \mathcal{O}\Big( \frac{T_U  +T_\rho}{\epsilon}\Big),$$
    where $T_\rho$ is the time required to construct $U_\rho$ and $T_U$ is the time required to construct $U$. 
\end{lemma}
\revise{In order to apply the above tool, we note that as we already have a block encoding of $A/s$ (Lemma~\ref{lemma: As}), it is straightforward to use it with Lemma~\ref{lemma: tensorproduct} to construct the block encoding of $M_1 \otimes \frac{A}{s} $, where $M_1$ is an arbitrary single-qubit gate (with known entries). Then we can use the above lemma to obtain an estimation of 
\begin{align}
\label{eq:TrMArho}
    \Tr\Big( \big( M_1 \otimes \frac{A}{s}  \big) \rho_k   \Big).
\end{align}
As $A/s$ is block-encoded with circuit complexity $\mathcal{O}\big( \log n + \log^{2.5} \frac{s}{\epsilon} \big)$, then the total complexity of estimating $  \Tr\Big( \big( M_1 \otimes \frac{A}{s}  \big) \rho_k   \Big)$ to accuracy $\epsilon$ is: 
\begin{align}
    \mathcal{O}\left( \frac{k }{\epsilon} \Big( \log\frac{1}{\epsilon} \Big)  \Big(\log(n) + \log^{2.5}\big(\frac{sk}{\epsilon}\big)  \Big)\right).
\end{align}}

\revise{Now we proceed to describe how to infer the desired value $\lambda = \bra{x_k}A\ket{x_k}$ from the above estimation. \\}

\revise{
\noindent\textbf{Procedure for finding $\lambda. $} First we write out the expression in Eqn.~(\ref{eq:TrMArho}):
\begin{align}
     & \Tr\Big( \big( M_1 \otimes \frac{A}{s}  \big) \rho_k   \Big) \\ &= \exp\Big(- \beta \big( 1-p_k \braket{x_k,\phi} \big)  \Big)^2 \Tr\big( M_1 \ket{0}\bra{0} \big) \Tr\big( \frac{A}{s}\ket{x_k}\bra{x_k} \big) + \Tr\big( M_1\ket{1}\bra{1}\big) \Tr\Big( \frac{A}{s} \sum_{j \neq 0}  \ket{\rm Garbage}_j  \bra{\rm Garbage}_j \Big). 
\end{align}
In the above, $\Tr\big( M_1\ket{0}\bra{0}\big)$ and $\Tr\big( M_1\ket{1}\bra{1}\big)$ are known exactly. }
\revise{If we repeat the same procedure, but with another single-qubit unitary $M_2$ instead of $M_1$, we obtain an estimation of:
\begin{align}
   & \Tr\Big( \big( M_2 \otimes \frac{A}{s}  \big) \rho_k   \Big) \\ &= \exp\Big(- \beta \big( 1-p_k \braket{x_k,\phi} \big)  \Big)^2 \Tr\big( M_2 \ket{0}\bra{0} \big) \Tr\big(\frac{A}{s}\ket{x_k}\bra{x_k} \big) + \Tr\big( M_2\ket{1}\bra{1}\big) \Tr\Big( \frac{A}{s} \sum_{j \neq 0}  \ket{\rm Garbage}_j  \bra{\rm Garbage}_j \Big). 
\end{align}}
\revise{It is convenient to define the following quantities.\begin{align}
    b_1 &= \Tr\Big( \big( M_1 \otimes \frac{A}{s}  \big) \rho_k   \Big),  \ \
    b_2 = \Tr\Big( \big( M_2 \otimes \frac{A}{s}  \big) \rho_k   \Big),  \\
    \lambda &= \Tr\big( \frac{A}{s}\ket{x_k}\bra{x_k} \big),  \ \
    \gamma = \Tr\big( \frac{A}{s} \sum_{j \neq 0}  \ket{\rm Garbage}_j  \bra{\rm Garbage}_j \big),  \\
    a_{11} &= \exp\Big(- \beta \big( 1-p_k \braket{x_k,\phi} \big)  \Big)^2 \Tr\big( M_1 \ket{0}\bra{0} \big), \ \
    a_{12}= \Tr\big( M_1\ket{1}\bra{1}\big), \\
    a_{21} &= \exp\Big(- \beta \big( 1-p_k \braket{x_k,\phi} \big)  \Big)^2 \Tr\big( M_2 \ket{0}\bra{0} \big), \ \
    a_{22} = \Tr\big( M_2\ket{1}\bra{1}\big).
\end{align}
By grouping the above pieces into a matrix and two vectors,
\begin{align}
    \mathscr{A} = \begin{pmatrix}
        a_{11} & a_{12} \\
        a_{21} & a_{22} 
    \end{pmatrix}, \xbf = \begin{pmatrix}
        \lambda \\
        \gamma
    \end{pmatrix}, \textbf{b} = \begin{pmatrix}
       s b_1 \\
       s b_2
    \end{pmatrix},
\end{align}
we arrive at the two by two linear system $\mathscr{A}\xbf = \textbf{b}$. Thus, solving them classically yields the value of $\lambda = \Tr\big( A \ket{x_k}\bra{x_k}\big)$, which is the desired quantity. The above system is only $2\times2$ matrix, so the classical processing is straightforward. However, we do not have an ideal linear system but rather an approximated one. More concretely, the values of $b_1$ and $b_2$ are obtained via Lemma~\ref{lemma: rall} via measurement, and we denote $\Tilde{\textbf{b}} \equiv (s \Tilde{b_1}, s\Tilde{b_2}  )^T$ as the approximated version of the ideal $\textbf{b} = (b_1,b_2)^T$. Similarly, the ideal entries $a_{11}$ and $ a_{21}$ are each estimated to accuracy $\epsilon$ via amplitude estimation, but the entries $a_{12},a_{22}$ are known exactly. Denote
\begin{align}
    \Tilde{\mathscr{A}} = \begin{pmatrix}
        \Tilde{a}_{11} & a_{12} \\
        \Tilde{a}_{21} & a_{22}
    \end{pmatrix}
\end{align}
as the matrix containing the estimated entries.  The approximated linear system is thus $\Tilde{\mathscr{A}} \Tilde{\xbf} = \Tilde{\textbf{b}}$.  }

\revise{Now we analyze the instability and error deviated from the ideal linear system. For our purpose, it is useful to utilize the following elementary result on the numerical linear algebra~\cite{trefethen2022numerical}:
\begin{theorem}
\label{thm: bound}
    Consider two linear equations $\mathscr{A} \xbf = \textbf{b}$ and $\Tilde{\mathscr{A}} \Tilde{x} = \Tilde{\bf b}$. If $A$ is non-singular and $|| \Tilde{\mathscr{A}} - \mathscr{A}  || \leq 1/||\Tilde{\mathscr{A}}^{-1}||$, then the following holds: 
    \begin{align*}
        \frac{||\Tilde{\xbf } - \xbf||}{||\xbf ||} \leq \frac{\kappa}{1-\kappa ||\Tilde{\mathscr{A}}-\mathscr{A} ||/||\mathscr{A}||} \cdot \Big( \frac{||\Tilde{\bf b}- \textbf{b}||}{|||\textbf{b}||} + \frac{ ||\Tilde{\mathscr{A}}- \mathscr{A} || }{|| \mathscr{A} ||}  \Big),
    \end{align*}
    where $||\cdot||$ refers to any norm measure and $\kappa$ is the conditional number of $A$. 
\end{theorem}} 
\revise{By choosing $||.||$ to be $l_2$ Euclidean norm, the application of the above theorem to our context is straightforward. The values of $b_1 = \Tr\Big( \big( M_1 \otimes \frac{A}{s}  \big) \rho_k   \Big)$ and $b_2 = \Tr\Big( \big( M_2 \otimes \frac{A}{s}  \big) \rho_k   \Big) $ are each estimated with an accuracy $\epsilon$. So, we have
\begin{align}
    ||\Tilde{\bf b}-\textbf{b}|| = \sqrt{ ( s\Tilde{b_1} -s b_1 )^2 + (s\Tilde{b_2}-sb_2)^2  }  = \sqrt{2} s\epsilon,
\end{align}
and
\begin{align}
    ||\Tilde{\mathscr{A}} - \mathscr{A}|| = \sqrt{ \big(\Tilde{a}_{11}-a_{11}  \big)^2 + \big(\Tilde{a}_{21}-a_{21}\big)^2   } = \sqrt{2} \epsilon.
\end{align}}  
\revise{Additionally, we have that 
\begin{align}
    ||\mathscr{A}||^2 &= a_{11}^2 +a_{12}^2 + a_{21}^2 + a_{22}^2  \\
    &=  \exp\Big(- \beta \big( 1-p_k \braket{x_k,\phi} \big)  \Big)^4 \Tr\big( M_1 \ket{0}\bra{0} \big)^2 \\&+ \Tr\big( M_1 \ket{1}\bra{1} \big)^2  + \exp\Big(- \beta \big( 1-p_k \braket{x_k,\phi} \big)  \Big)^4 \Tr\big( M_2 \ket{0}\bra{0} \big)^2  + \Tr\big( M_2 \ket{1}\bra{1} \big)^2 \\
    &= \mathcal{O}(1),
\end{align}
which is due to the fact that $0.5 \leq \exp\big(-\beta (1-x)\big) \leq 1$ for $0\leq x\leq 1$. So we have
$$0.25 \Tr\big( M_1 \ket{0}\bra{0} \big)^2\leq  exp\Big(- \beta \big( 1-p_k \braket{x_k,\phi} \big)  \Big)^4 \Tr\big( M_1 \ket{0}\bra{0} \big)^2 \leq  \Tr\big( M_1 \ket{1}\bra{1} \big)^2, $$
and similarly 
$$0.25\Tr\big( M_2 \ket{0}\bra{0} \big)^2  \leq  exp\Big(- \beta \big( 1-p_k \braket{x_k,\phi} \big)  \Big)^4 \Tr\big( M_2 \ket{0}\bra{0} \big)^2 \leq  \Tr\big( M_2 \ket{1}\bra{1} \big)^2. $$
These lead to the consequence that
\begin{align}
    ||\mathscr{A}||^2 \leq \Tr\big( M_1 \ket{0}\bra{0} \big)^2+ \Tr\big( M_1 \ket{1}\bra{1} \big)^2 + \Tr\big( M_2 \ket{0}\bra{0} \big)^2 + \Tr\big( M_2 \ket{1}\bra{1} \big)^2  = 1
\end{align}
and  
\begin{align}
    0.25 \Big( \Tr\big( M_1 \ket{0}\bra{0} \big)^2 + \Tr\big( M_2 \ket{0}\bra{0} \big)^2 \Big) + \Tr\big( M_1 \ket{1}\bra{1} \big)^2 + \Tr\big( M_2 \ket{1}\bra{1} \big)^2 \leq ||\mathscr{A}||^2 \\
    \longrightarrow 0.25 \Big( \Tr\big( M_1 \ket{0}\bra{0} \big)^2+ \Tr\big( M_1 \ket{1}\bra{1} \big)^2 + \Tr\big( M_2 \ket{0}\bra{0} \big)^2 + \Tr\big( M_2 \ket{1}\bra{1} \big)^2  \Big) \leq ||\mathscr{A}||^2   \\
    \longrightarrow 0.5 \leq ||\mathscr{A}||^2 ,
\end{align}
where we also make use of the unitary property of $M_1,M_2$, which implies that $\Tr\big( M_1 \ket{0}\bra{0} \big)^2+ \Tr\big( M_1 \ket{1}\bra{1} \big)^2 =1  $ and $ \Tr\big( M_2 \ket{0}\bra{0} \big)^2 + \Tr\big( M_2 \ket{1}\bra{1} \big)^2  = 1$. }
\revise{Summing up, we have: 
\begin{align}
     \frac{||\Tilde{\xbf } - \xbf||}{||\xbf ||} &=\frac{\kappa}{1-\kappa ||\Tilde{\mathscr{A}}-\mathscr{A} ||/||\mathscr{A}||} \cdot \Big( \frac{||\Tilde{\bf b}- \textbf{b}||}{|||\textbf{b}||} + \frac{ ||\Tilde{\mathscr{A}}- \mathscr{A} || }{|| \mathscr{A} ||}  \Big) \\
     &\leq \frac{\kappa}{(1- \kappa \sqrt{2}\epsilon)}  \cdot \Big( \frac{\sqrt{2}s \epsilon}{||\textbf{b}||} + 2\epsilon  \Big) \\
      &=  \frac{\kappa}{(1-\kappa \sqrt{2}\epsilon)}  \Big(  \frac{\sqrt{2} s}{||\textbf{b}||} + 2 \Big) \epsilon\\
     &\longrightarrow ||\Tilde{\xbf } - \xbf|| \leq \frac{\kappa}{(1-\kappa \sqrt{2}\epsilon)} \Big( \frac{\sqrt{2}||\xbf||s  }{||\textbf{b}||    }  + 2 ||\xbf|| \Big) \epsilon.
\end{align}
Since $\Tilde{\mathscr{A}}\xbf = \Tilde{\textbf{b}}$, so $\xbf = \Tilde{\mathscr{A}}^{-1} \Tilde{\textbf{b}} $, and thus:
\begin{align}
    ||\xbf|| = || \Tilde{\mathscr{A}}^{-1} \textbf{b}|| \leq ||\Tilde{\mathscr{A} }^{-1}|| \cdot ||\textbf{b}|| \\
    \longrightarrow \frac{||x||}{||\textbf{b}||} \leq ||\Tilde{\mathscr{A}}^{-1}|| = \kappa ||\mathscr{A}||.
\end{align}
As we have pointed out before that $||\mathscr{A}|| =\mathcal{O}(1)$, we further have that  $||\textbf{x}|| = \sqrt{\lambda^2 + \gamma^2} \leq \sqrt{1+1} = \sqrt{2}$ and thus 
\begin{align}
    ||\Tilde{\xbf } - \xbf||  \leq \frac{\kappa}{(1-\kappa \sqrt{2}\epsilon)} \Big(  \sqrt{2} s \mathcal{O}(1) + 2\sqrt{2}  \Big) \epsilon.
\end{align}
For a sufficiently small $\epsilon$, for example, $\epsilon \leq \frac{1}{2\sqrt{2}\kappa}$, then $ 1-\kappa \sqrt{2}\epsilon \geq \frac{1}{2}$, so  we have
\begin{align}
    ||\Tilde{\xbf } - \xbf|| \leq  2\kappa \big( \sqrt{2}\mathcal{O}(1)s + 2\sqrt{2} \big) \epsilon \\
    \longrightarrow ||\Tilde{\xbf } - \xbf|| \leq C\kappa  \epsilon,
\end{align}
where we have defined $C \equiv 2(\sqrt{2}s\mathcal{O}(1) + 2\sqrt{2})$ and $\xbf \equiv (\lambda,\gamma)^T$, and $\Tilde{\xbf}$ is the solution of the approximated linear system. Suppose that $\Tilde{\xbf }= (\Tilde{\lambda}, \Tilde{\gamma})^T$  contains the estimation of the ideal $\lambda,\gamma$, we can calculate the difference:
\begin{align}
    |\Tilde{\lambda}-\lambda| &\leq \sqrt{|\Tilde{\lambda}-\lambda|^2+ | \Tilde{\gamma}-\gamma|^2}\\
    &= ||\Tilde{\xbf}-\xbf|| \\
    &\leq  C\kappa \epsilon.
\end{align}
At this point, we have obtained a full procedure that allows us to obtain a quantity $\Tilde{\lambda}$ such that $|\Tilde{\lambda}-\lambda| =C\kappa \epsilon $ (we note that $\kappa$ is conditional number of $\Tilde{\mathscr{A}}$), using a quantum circuit of depth:
\begin{align}
    \mathcal{O}\left( \frac{k}{\epsilon}\log\frac{1}{\epsilon}  \Big(\log(n) + \log^{2.5}\big(\frac{sk}{\epsilon}\big)  \Big)\right). 
\end{align} }
\revise{The above bound clearly implies that the conditional number of the system $\mathscr{A}$ is a crucial quantity. %For convenience, we recall that:
%\begin{align}
%   \mathscr{A} = \begin{pmatrix}
%        \exp\Big(- \beta \big( 1-p_k \braket{x_k,\phi} \big)  \Big)^2 \Tr\big( M_1 %\ket{0}\bra{0} \big) & \Tr\big( M_1\ket{1}\bra{1}\big) \\
%        \exp\Big(- \beta \big( 1-p_k \braket{x_k,\phi} \big)  \Big)^2 \Tr\big( M_2 %\ket{0}\bra{0} \big) & \Tr\big( M_2 \ket{1}\bra{1}\big)
%    \end{pmatrix}
%\end{align} }\textcolor{blue}{[It would be useful to draw vertical and horizontal lines to separate the four elements in the matrix for visual convenience.]
}
\revise{ All entries of $\mathscr{A}$ has magnitude of $\mathcal{O}(1)$, and we have the freedom to choose $M_1$ and $M_2$, we can make sure that the conditional number of $\mathscr{A}$ is $O(1)$. We recall that previously, at one step, we performed the transformation of the block-encoded operator $p_k \braket{x_k,\phi} \ket{x_k}\bra{\phi} \longrightarrow \exp\Big( -\beta\big( 1- p_k \braket{x_k,\phi}  \big)  \Big) $, which was somewhat unintuitive, as one may wonder why we choose the function $\exp \big( -\beta(1-x) \big)$. The answer is to avoid the potentially exponential large value of $\kappa$ due to small $p_k$. A justification for this choice is based on the property that this particular function dramatically amplifies the value of input $x$ (see the following figure). }
\revise{
\begin{figure}[h!]
    \centering
    \begin{tikzpicture}[scale = 2.0]
        \draw[->] (-1.2, 0) -- (1.2, 0) node[right] {$x$};
    \draw[->] (0, -0.5) -- (0, 2) node[above] {$y$};
    \draw[domain = -1:1, smooth, variable = \x, blue, thick] plot( {\x}, { exp(-0.01*(1-\x)} );
    \filldraw[black] (1, 0) circle (1pt)  ;
    \filldraw[black] (-1, 0) circle (1pt)  ;
    \node at (1, -0.2) [black] {(1,0)} ;
    \node at (-1, -0.2) [black] {(-1,0)  }; 
    \draw[dashed]  (1,0) -- (1,1) ;
    \draw[dashed] (-1,0) -- (-1,1) ;
    \end{tikzpicture}
    \caption{The plot of function $\exp\big(-\beta (1-x)\big)$ for $\beta = 0.01$, which essentially amplifies the value of $x$ above some threshold, for example, above the line $y=0.5$.  }
    \label{fig: exp}
\end{figure} }
\revise{
If we don't apply such transformation and proceed with the algorithm with the block-encoded operator $ p_k \braket{x_k,\phi} \ket{x_k}\bra{\phi}$, then eventually the matrix $\Tilde{A}$ would become:
\begin{align}
    \begin{pmatrix}
        p_k \Tr\big( M_1 \ket{0}\bra{0} \big) & \Tr\big( M_1\ket{1}\bra{1}\big)  \\
        p_k \Tr\big( M_2 \ket{0}\bra{0} \big) & \Tr\big( M_2 \ket{1}\bra{1}\big)
    \end{pmatrix},
\end{align}}
\noindent\revise{and its conditional number could be exponentially large with respect to $k$. To show 
this, we observe that the determinant of the above matrix is $p_k \Tr\big( M_1 
\ket{0}\bra{0} \big)\Tr\big( M_2 \ket{1}\bra{1}\big) -  p_k \Tr\big( M_2 \ket{0}\bra{0} \big) \Tr\big( M_1\ket{1}\bra{1}\big)  \sim p_k$. Furthermore, we have defined $p_k = 
||x_k||^2/s^k$, which clearly indicates that the determinant is exponentially small in 
$k$. This means that the smaller singular value/eigenvalue is exponentially small in $k$, and thus the conditional number, which is the ratio between the largest and smallest singular value/eigenvalue, is exponentially large in $k$. Such an exponential dependence inflates the previously derived bound $||\Tilde{\xbf} - 
\xbf|| \leq 2\kappa ||\xbf|| \epsilon$, which means that we need to choose $\epsilon$ to 
be exponentially small, thus resulting in high complexity. So the choice of function 
$\exp\big(- \beta (1-x) \big)$ with $\beta = 0.01$ can avoid such potential scenario, regardless of the input, the output will be very sufficiently large, e.g., at least greater than $1/2$, 
as indicated in Figure \ref{fig: exp}.  
}

\smallskip
\revise{In brief, we have successfully constructed a quantum algorithm for performing the power method with $k$ iterations, with a careful stability analysis. The output of the aforementioned procedure is a quantity $\Tilde{\lambda}$ that is $\epsilon$-closed to $\lambda = \bra{x_k}A\ket{x_k}$, i.e., $|\Tilde{\lambda}- \lambda| \leq C\kappa \epsilon$ where $\kappa =\mathcal{O}(1)$. The quantum circuit complexity is:
\begin{align}
    \mathcal{O}\left( \frac{k}{\epsilon}\log\frac{1}{\epsilon}  \Big(\log(n) + \log^{2.5}\big(\frac{sk}{\epsilon}\big)  \Big)\right). 
\end{align}
We remark further that $\lambda$ is still an approximation to the true largest eigenvalue $\lambda_{\max}$ of $A$. As we mentioned from the beginning, to get $\lambda$ to be $\epsilon$ close to $\lambda_{\max}$, we need $k=\mathcal{O}\Big(  \log(n)+\log \frac{1}{\epsilon}\Big)$. As the error adds up linearly, we have that the total error is $C\kappa \epsilon + \epsilon = (C\kappa+1)\epsilon$.  Thus, in order to estimate the largest eigenvalue to additive error $\delta$, we need to set $(C\kappa+1)\epsilon=\delta \longrightarrow \epsilon = \frac{\delta}{C\kappa+1}$, where we recall that $C= 2\sqrt{2}(\mathcal{O}(1)s +2) = \mathcal{O}(s)$. Thus, we arrive at the main result as the following theorem:
\begin{theorem}[Improved Quantum Power Method]
\label{thm: imqpm}
    Given oracle access to entries of a Hermitian matrix, $s$-sparse $A$. Let $\ket{x_0}$ be some initial state prepared by circuit of $\mathcal{O}(1)$ depth. Let $x_k = A^k \ket{x_0}$ and $\ket{x_k} = x_k/||x_k||$ be its normalized quantum state. Then the quantity $\lambda = \bra{x_k}A\ket{x_k} $
can be estimated up to accuracy $\epsilon$ in complexity
\begin{align*}
    \mathcal{O}\left( \frac{k}{\epsilon}\log\frac{1}{\epsilon}  \Big(\log(n) + \log^{2.5} \frac{sk}{\epsilon}  \Big)\right)
\end{align*}
As detailed in the Appendix \ref{sec: powermethoditeration}, by choosing $k = \mathcal{O}\left(  \frac{1}{\Delta} \log\Big( \frac{2 }{ \delta\cos^2(\theta_0)}\Big)  \right) $ (where $\cos^2(\theta_0)$ being the squared overlaps between $\ket{x_0}$ and the eigenvector corresponding to the largest eigenvalue of $A$, $\Delta$ is the gap between two largest eigenvalues) and $\epsilon = \delta/\mathcal{O}(s)$, the largest eigenvalue of $A$ in magnitude $\lambda_{\max}$ can be estimated with additive accuracy $\delta$ in complexity
    \begin{align*}
        \mathcal{O} \left(  \frac{s}{\delta} \log\big( \frac{s}{\delta}\big) \frac{1}{\Delta} \log\Big( \frac{2 }{ \delta\cos^2(\theta_0)}\Big)   \Big(\log(n) + \log^{2.5}\big(\frac{s^2k}{\delta}\big)  \Big)  \right).
    \end{align*}
\end{theorem} }
\revise{For the purpose of comparison, we recapitulate the running time of the original proposal on quantum power method ~\cite{nghiem2022quantum}:
\begin{lemma}[\cite{nghiem2022quantum}]
    Given oracle access to entries of a Hermitian, $s$-sparsee matrix $A$ and some initial state $\ket{x_0}$. Le  $x_k = A^k \ket{x_0}$ and $\ket{x_k}$ be its normalized version, then the largest eigenvalue (in magnitude) of $A$, $\lambda_{\max}$ can be estimated up to an additive error $\delta$ in time
$$ \mathcal{O}\Big(   \frac{\log(n) \sqrt{n} s \kappa^2     }{\delta^4}\Big), $$
where $\kappa$ is the conditional number of $A$.
\label{lemma: qpm}
\end{lemma}
We point out specific aspects of improvement. The scaling on error tolerance $\delta$ is reduced from $\frac{1}{\delta^4}$ to $\frac{1}{\delta} \rm polylog \frac{1}{\delta}$, which is almost a power-of-4 improvement. The dependence on dimension is also reduced from quadratic to polylogarithmic. Additionally, the dependence on conditional number $\kappa$ is removed. Evidently, the two complexities differ significantly, which clearly shows the major enhancement of our proposal. As emphasized in the overview section~\ref{sec: overviewquantumpowermethod}, the bottleneck imposed in~\cite{nghiem2022quantum} is the measurement required to obtain the desired state $\ket{x_k}$. As shown in~\cite{nghiem2022quantum}, the success probability turns out to be exponentially small in $k$, which resulted in many repetitions in order to obtain $\ket{x_k}$. Our method does not rely on intermediate measurements, thus avoiding a costly step. 
}

\section{Improving the Gradient Descent Method}
\label{sec: gradientdescent}
\revise{In this section}, we proceed to describe the \revise{new frameworks for performing gradient descent in the quantum setting, providing an alternate perspective to the original proposal~\cite{rebentrost2019quantum}}. We shall see that \revise{leveraging the block encoding technique and related arithmetic operations} can remove major complicated steps, providing a relatively simpler yet \revise{significantly more enhanced} method. \revise{As we will show}, \revise{the dependence on the inverse of error tolerance is polylogarithmic, which is a superpolynomial improvement to~\cite{rebentrost2019quantum}}. For simplicity, we set a convention that $||A|| < 1$, where $||\cdot||$ refers to the matrix norm, as it is always achievable by trivial scaling, e.g., by considering the new matrix $A' = A/|A_{max}|$ instead, where $A_{max}$ is the maximaml element in $A$. As a result, also pointed out in~\cite{rebentrost2019quantum}, the norm of the gradient operator $D$, denoted as $||D||$, is upper bounded by $p$ for arbitrary input during the iteration. Such an assumption does not incur any systematic issue but is merely made for simplification. Furthermore, we assume that the oracle access to each term in the summation of $M_D$ (Eqn. \ref{eqn: md}) is available, similar to that of~\cite{rebentrost2019quantum}.  \\

\subsection{First Version}
\label{sec: firstversion}
In the first version, our problem statement is somewhat an alternate to that of~\cite{rebentrost2019quantum}. First, we relax the spherical constraint condition and treat the minimization problem in its full generality. Next, for each iteration step $t$, instead of producing the exact quantum state $\ket{x_t}$, we aim to produce the block encoding of a matrix that is proportional to $x_t (x_t)^T$. (Note that we will deal with real vectors $x_t$'s, so the Hermitian conjugate produces the same effect as the transpose, and we use transpose for convenience.) The reason for such a representation is more or less inspired by our main tool, i.e., the unitary block encoding~\cite{gilyen2019quantum, low2017optimal, low2019hamiltonian}. The framework allows a highly flexible and universal way to manipulate an arbitrary matrix by encoding it into a unitary matrix. Therefore, if we can produce the desired state, or more generally, the density matrix in such a unitary block encoding representation, then, in principle, we can extract useful information from such a representation. A marked difference between this alternative problem compared to that of~\cite{rebentrost2019quantum} is because we do not require the spherical constraint, e.g., at each iteration step $t$, and the output is a quantum state (normalized vector) $\ket{x_t}$. Therefore, we do not employ the intermediate measurement and, hence, relax a major step that could contribute substantially to the running time due to the finite probability of success in measurement.   

In the above formulation of $A$ and $D$, we see that there are tensor product structures in their formulation. As such, the first recipe we will need is the production of the unitary block encoding of the tensor product of some operators, given the unitary block encoding of each operator, respectively. For convenience, we recall the following tool from~\cite{camps2020approximate}.
\begin{lemma}[\cite{camps2020approximate}]
    Given unitary block encoding $\{U_i\}_{i=1}^m$ of multiple operators $\{M_i\}_{i=1}^m$ (assumed to be exact encoding). Then there is a procedure that produces a unitary block encoding operator of $\bigotimes_{i=1}^m M_i$, which requires a single use of each $\{U_i\}_{i=1}^m$ and $\mathcal{O}(1)$ SWAP gates. 
\end{lemma}

Since each term in the summation of $M_D$ (Eqn.~\ref{eqn: md}), e.g., $M_j$, has entries given via the oracle access, the same assumption as given in~\cite{rebentrost2019quantum}, and the preparation of $\epsilon$-approximation unitary encoding of $M_j/s$ is achievable due to Lemma~\ref{lemma: As}, where $s$ is the sparsity of $A$ (see e.g., chapter 27 of~\cite{childs2017lecture}). Furthermore, in~\cite{gilyen2019quantum}, the authors showed the following: 
\begin{lemma}
    Given unitary block encoding of multiple operators $\{M_i\}_{i=1}^m$. Then there is a procedure that produces a unitary block encoding operator of $\sum_{i=1}^m \pm M_i/m $ with a time complexity $\mathcal{O}(m)$, 
    \label{lemma: sumencoding}
\end{lemma}
where the $\pm$ sign in the above lemma indicates that it can be either addition or subtraction. Therefore, the $\epsilon$-approximation block encoding of $M_D/( sp )$ is easily obtained in time $\mathcal{O}(p^2\log(n) + p \log^{2.5}(1/\epsilon) ))$. 

It is also straightforward to see that  Lemma~\ref{lemma: tensorproduct} allows us to obtain the unitary block encoding of $(x x^T)^{\otimes (p-1)} \otimes \mathbb{I}$ given the unitary block encoding of $x x^T$, since the block encoding of $\mathbb{I}$ is simple. For example,  the matrix corresponding to the controlled-Z gate  
\begin{align*}
  c-Z =  \begin{pmatrix}
        1 & 0 & 0 & 0 \\
        0 & 1 & 0 & 0\\
        0 & 0 & 1 & 0\\
        0 & 0 & 0 & -1
    \end{pmatrix}
\end{align*}
is a block encoding of the $2\times 2$ identity using two qubits (if we focus on the top left corner of the $2\times2$ block structure). Therefore, the tensor product $c-Z \otimes \mathbb{I}_d$ (for an arbitrary dimension $d$) will produce a trivial block encoding of the identity matrix of an arbitrary dimension. 

In order to produce the block encoding of the gradient operator $D$, we can use the following procedure. With the block encodings of $M_D/(ps)$ and $(x x^T)^{\otimes (p-1)} \otimes \mathbb{I}$, it is straightforward to use Lemma~\ref{lemma: product} to prepare the block encoding of the operator 
$$ (x x^T)^{\otimes (p-1)} \otimes \mathbb{I} \,\,\cdot\frac{M_D}{ps} \,\cdot (x x^T)^{\otimes (p-1)} \otimes \mathbb{I}.  $$
Using Eq.~(\ref{eq:MD}), we can straightforwardly show that 
\begin{align}
    (x x^T)^{\otimes (p-1)} \otimes \mathbb{I} \,\,\cdot\frac{M_D}{ps}\,\cdot (x x^T)^{\otimes (p-1)} \otimes \mathbb{I} 
    =  (x x^T)^{\otimes p-1} \otimes \frac{D(x)}{ps}.
\end{align}
In the appendix, we show the following lemma:
\begin{lemma}
\label{lemma: 11}
    Given the block encoding of $(x x^T)^{\otimes p-1} \otimes \frac{D(x)}{ps}$, then it is possible to obtain the block encoding of $D(x)/ps$ in time $\mathcal{O}( \gamma^{2(p-1)} ps )$, where $\gamma$ is some constant.
\end{lemma}

The factor $\gamma$ in the above lemma will be derived in appendix~\ref{sec: proof11}, and it will be shown there that $\gamma$ is upper bounded by either heuristic method or by choosing the learning rate $\eta$ in the gradient descent sufficiently small. Now, we are ready to outline our improved quantum gradient descent in detail. \\

\noindent\textbf{Improved Quantum Gradient Descent} \\

$\bullet$ The first step in the gradient descent method is the creation of a random real vector $x_0$. Without loss of generality, we can take $x_0$ to be generated by some known circuit (therefore, it has a unit norm). We need to prepare a unitary encoding of $x_0 x_0^T$. To do so, we first note the decomposition of $x_0$, or $\ket{x_0}$ in the standard computational basis as:
\begin{align}
    \ket{x_o} = \sum_{i=1}^n x_0^i \ket{i}.
\end{align}
If we then add a second register initialized in $\ket{0}$ and trace it out, then we trivially obtain: 
\begin{align}
    \rho = \sum_{i,j=1}^n x_0^i x_0^j \ket{i}\bra{j},
\end{align}
 which is exactly $\ket{x_0}\bra{x_0}$. It is thus straightforward to apply Lemma~\ref{lemma: improveddme} to prepare the unitary encoding of $x_0 x_0^\dagger$. The generation of $\ket{x_0}$ can be done with an arbitrary low-depth circuit, e.g., one with a few rotation gates. Therefore, it is safe to assume that the time required is $\mathcal{O}(1)$, as we will choose some random vector $x_0$ that can be generated with a short-depth circuit. Therefore, the complexity for preparing the block encoding of $\rho$ as above is $\mathcal{O}(\log(n))$. \\

$\bullet$ Given the randomly chosen state $x_0$, the previous step yields the unitary encoding of $x_0 x_0^T$.  Then we need to perform the following iteratively: 
\begin{align}
    x_{t+1} = x_t  - \eta \bigtriangledown f(x_t).
\end{align}
In the density matrix representation, the above formulation can be rewritten as:
\begin{align}
    x_{t+1} (x_{t+1})^T &=  (x_t  - \eta \bigtriangledown f(x_t))( x_t  - \eta \bigtriangledown f(x_t)  )^\dagger \\
    &= (I - \eta D(x_t)) x_t x_t^T (I  - \eta D(x_t)^\dagger) \\
    &= (I - \eta D(x_t)) x_t x_t^T (I- \eta D(x_t)),
    \label{eqn: 42}
\end{align}
where in the last line, we have used the hermitian property of the gradient operator $D$. Previously, we have shown that we can prepare the unitary encoding of $D(x_t)/(p s )$ given the unitary block encoding of $x_t x_t^T$. Note that the factor $p s$ is non-trivial, and we only need to make a corresponding adjustment. In this case, we need to obtain the unitary block encoding of $(I - D(x_t))/(  p s) $, which requires the unitary block encoding of $I / (p s)$, which has dimension $n \times n$. We consider the following procedure to achieve the unitary block encoding of $I / (ps)$. Recall that the RY rotational gate has the matrix representation:
\begin{align}
   R_Y(\theta) = \begin{pmatrix}
        \cos(\theta/2) & -\sin(\theta/2) \\
        \sin(\theta/2) & \cos(\theta/2) 
    \end{pmatrix}.
\end{align}
If one choose $\theta$ such that 
$\cos( \theta/2) ={1}/{ p s}$, 
then $R_Y(\theta)$ has $\frac{1}{ps} $ on the diagonals. The tensor product of $R_y(\theta)$ with $\mathbb{I}_2$ (where $\mathbb{I}_2$ is the identity matrix of size $2 \times 2$), i.e.,
%\begin{align*}
    $R_Y(\theta/2) \otimes \mathbb{I}_2$,
%\end{align*}
contains the following matrix
\begin{align*}
    \begin{pmatrix}
        \frac{1}{ sp}  & 0 \\
        0 & \frac{1}{ sp}  
    \end{pmatrix}
\end{align*}
on the top-left corner. Therefore, $R_Y(\theta) \otimes \mathbb{I}_2$ is exactly the unitary block encoding of the above matrix. Keep repeating the process, i.e., the tensor product with $\mathbb{I}_2$, we will obtain the unitary block encoding of an $n \times n$ matrix that contains 
$\frac{1}{ s p }$
on the diagonals,  which is nothing but $I /(sp) $ (where we have omitted the dimension $n$ in the subscript). The same procedure as above also allows us to build the unitary block encoding of $diag(\eta)$, whose nonzero entries are all $\eta$ on the diagonals. Then, we can use Lemma~\ref{lemma: product} to construct the unitary block encoding of $\eta D(x_t)/(ps)$.    \\

Given the block encoding of $I /(s  p) $ as outlined above (note that we are only concerned with the top left corner block) plus the unitary block encoding of $x_0 x_0^T$,  Lemma~\ref{lemma: product} allows us to construct the unitary block encoding of $U_1 \equiv x_0 x_0^T /(sp ) $. Similarly, the block encoding of, say, $\eta D(x_0) /(sp )$ and of $x_0x_0^T$ allows us to construct the block encoding of their product, $U_2 \equiv \eta D(x_0) x_0 x_0^T /(s p)$  and the reverse of their product, i.e., $U_3 \equiv x_0 x_0^T \eta D(x_0) /(s p)$. Furthermore, Lemma~\ref{lemma: product} combines the block encoding of $\eta D(x_0) x_0 x_0^T /(s  p)$ and $\eta D(x_0) /(sp )$ to yield the block encoding of  
\begin{align}
    U_4 \equiv \frac{ \eta D(x_0) }{(sp) } x_0 x_0^T \frac{ \eta D(x_0) }{(s p) }.
    \label{eqn: U}
\end{align}

Note that there is an unwanted factor $(sp)$, which might accumulate to a larger factor as we continue to iterate the process. Fortunately, this issue has been handled in~\cite{gilyen2019quantum}, where the authors show that by using a technique called preamplification, it is possible to get rid of such a factor using $\mathcal{O}( sp )$ corresponding block-encoding operators, e.g., $U_2$. Given $U_1,U_2,U_3, U_4$ (we already assume that the unwanted factor is removed), Lemma~\ref{lemma: sumencoding} allows us to construct the block encoding of 
$$ \frac{U_1 - U_2 - U_3 + U_4}{4}. $$
It is then straightforward to verify the following property:
\begin{align}
    x_0 x_0^T - \eta D(x_0) x_0 x_0^T - x_0 x_0^T \eta D(x_0) + \eta D(x_0) x_0 x_0^T \eta D(x_0) &= (I-  \eta D(x_0) x_0 x_0^T (I - \eta D(x_0)) \\
    &= x_1 x_1^T,
\end{align}
where the last line comes from Eqn.~(\ref{eqn: 42}). Therefore, the matrix defined in Eqn.~(\ref{eqn: U}) is exactly
${x_1 x_1^T}/{4}$.
We can iteratively proceed with the above procedure to obtain the desired operator for a fixed time $T$. We note that the factor $4$ from the above could be removed using preamplitication method~\cite{gilyen2019quantum}. In the appendix, we would show that if one carefully chooses the initial operator $x_0 x_0^T$ with bounded norm, then the norm of operator $x_t x_t^T$ is less than unity at any $t$-th iteration step.  \\

As a summary, our algorithm begins with preparing unitary encoding of $M_j$ (see~\ref{eqn: md}). Each unitary encoding of $M_j$ takes $p\log(n)$ time to prepare since a SWAP operation between two $p\log(n)$-qubit systems is required to do this (see further~\cite{low2017optimal, gilyen2019quantum}). The preparation of $M_D$ in Eqn.~(\ref{eqn: md}), by Lemma~\ref{lemma: sumencoding}, will incur $\mathcal{O}(p)$ operations of $M_j$'s. The preparation of the unitary block encoding for the gradient operator $D/ps$ with
\begin{align}
    D = tr_{1,2,...,p-1} ( (x x^{T})^{\otimes (p-1)} \otimes \mathbb{I}) M_D
\end{align}
takes $\mathcal{O}(\alpha^{2p-2} )$ operations due to the step in taking the  trace. As we can see, such a time is negligible compared to the time required by $M_D$, as the summation requires $\mathcal{O}( p \log n)$ operations. Lastly, we use the preamplification method to remove the undesired subnormalization factor, which takes further $\mathcal{O}( sp)$ time. Now, we summarize the main result of this section before comparing it with the original version in~\cite{rebentrost2019quantum}. 

\begin{theorem}
\label{thm: main1}
    Given a function $f: \mathbb{R}^n \rightarrow \mathbb{R}$ with algebraic form as defined in Eqn.~(\ref{eqn: algebraicform}). For a fixed iteration number $T$, there exists a quantum algorithm that produces a unitary $U_T$ that encodes an $\epsilon$-approximation of the  operator 
    $ x_T x_T^T $.
    Define 
    $\beta \equiv \alpha^{2p-2} ps$,
   where $\alpha$ is some bounded constant. The running time of the quantum algorithm is
    $$ \mathcal{O}\Big(  ( \log(n) + \log^{2.5}\frac{1}{\epsilon})  \beta \frac{\beta^{T+1}-1}{\beta-1} \Big).  $$
\end{theorem}

Given the output in Thm~\ref{thm: main1}, one can perform arbitrary subsequent processing on the block-encoded matrix, which contains $x_T x_T^T$ without using measurement. As to what kind of useful operation depends on the kind of application. 
To compare the complexity with the original algorithm, we recall that from~\cite{rebentrost2019quantum}, the time complexity for producing the final state $\ket{x_T}$ up to accuracy $\epsilon$ is 
$$ \mathcal{O}\Big( \big(\frac{p}{\epsilon} \big)^{3T} \big( \frac{p s_A}{ \epsilon } \big)^T \log(n)   \Big).  $$
The seeming advantage of the above algorithm relative to that of~\cite{rebentrost2019quantum} is that the running time is much more efficient with respect to error tolerance $\epsilon$, and polynomial improvement over $T$. However, as we have commented previously, the goal of our method is a bit different from that of~\cite{rebentrost2019quantum}, but what we will do next is to introduce another algorithm closer to that of~\cite{rebentrost2019quantum}, where we expect that the unitary block encoding framework can also provide some enhancement.

\subsection{Second Version}
\label{sec: secondversion}
In this setting, we directly improve upon the method in~\cite{rebentrost2019quantum}, which means that we impose the spherical constraint. Our goal now is to produce the quantum state $\ket{x_t}$ that corresponds to the temporal solution after $t$-th iteration step. Similar to~\cite{rebentrost2019quantum}, suppose at step $t$, we are presented with multiple copies of $\ket{x_t}$, which can be the output of the previous step. The new solution is then updated per the following formula:
\begin{align}
    x_{t+1} = \ket{x_t} - \eta D(\ket{x_t}\bra{x_t}) \ket{x_t}.
    \label{eqn: 49}
\end{align}
Then, of course, we need to obtain the quantum state $\ket{x_{t+1}}$ that corresponds to the (un-normalized) vector $x_{t+1}$. We note that in this condition, the formula for the gradient operator (Eqn.~\ref{eqn: D}) needs to take into account such a normalization issue as well because we will be working with a quantum state instead of a general vector. 

What we do now is to construct the unitary block encoding of $\rho_t \equiv \ket{x_t}\bra{x_t}$. Since we are only presented with copies of $\rho_t$, we cannot apply Lemma~\ref{lemma: improveddme}, as we do not have the means to prepare the purification of $\rho_t$. We first need the so-called density matrix exponentiation technique that was proposed in~\cite{lloyd2014quantum}. 

\begin{lemma}[\cite{lloyd2014quantum}]
\label{lemma: dme}
   Given multiple copies of $\rho$, the unitary $\exp(-i \rho t)$ can be simulated up to error $\epsilon$ in time 
   $ \mathcal{O}\Big( t^2 \epsilon^{-1}  \Big)$, using $\mathcal{O}(t^2 \epsilon^{-1})$ copies of $\rho$.  
\end{lemma}
The following result from~\cite{gilyen2019quantum} is also needed.
\begin{lemma}[Corollary 71 in \cite{gilyen2019quantum}]
\label{lemma: logU}
    Given the unitary $\exp(-i \rho)$, it is possible to construct a $\delta$-close unitary block encoding of $ \pi \rho/ 4$ using
    $ \mathcal{O}\Big( \log\big( \frac{1}{\delta} \big)   \Big)$
    controlled versions of $\exp(-i \rho )$ and its inverse plus a single ancilla qubit. 
\end{lemma}
We note that the technique in Lemma \ref{lemma: dme} also allows us to construct a controlled version of $\exp(-i\rho t)$. Given the unitary $\exp(-i \rho t)$ plus its controlled version, if we set $t = 1$, then it is possible to construct an approximation of the unitary block encoding of $\pi\rho/4$ by applying the above lemma. This step incurs a quantum circuit of depth $\mathcal{O}\Big( \frac{1}{\epsilon} \log \frac{1}{\epsilon}   \Big)$ and uses $\mathcal{O}\Big( \frac{1}{\epsilon} \log \frac{1}{\epsilon}   \Big)$ copies of $\rho$. 

Now, we have enough preparation to solve our problem. By Lemma~\ref{lemma: sumencoding}, the unitary encoding of $(I - \eta M_D)/p s $ can be constructed, as each of the terms in the subtraction can be block encoded. Lemma~\ref{lemma: logU} yields the $\delta$-close block encoding of $\pi \rho_t/ 4$, which can be used to construct the $p\delta$-close block encoding of $(\pi \rho_t/ 4 )^{\otimes p}$. Then Lemma~\ref{lemma: product} allows us to construct the (approximated) unitary block encoding of their product, i.e., 
\begin{align}
     \frac{I -\eta M_D }{sp} \cdot (\frac{\pi \rho_t}{4 } )^{\otimes p}.
\end{align}

We use the approximated block encoding of $(\pi \rho_t/ 4)^{\otimes p-1}\otimes \mathbb{I}$ (note that the last term is the identity matrix) plus Lemma~\ref{lemma: product} to construct the block encoding $U_t$ of 
$$P_t \equiv  \big(\frac{\pi\rho_t}{4}\big)^{\otimes p-1}\otimes \mathbb{I} \cdot \frac{(I - \eta M_D) }{sp} \cdot (\frac{ \pi \rho_t}{4 } )^{\otimes p}. $$

Now, we pay attention to the numerator. As $\rho_t$ is a projector, $\rho_t \cdot \rho_t = \rho_t$, we have: 
\begin{align}
  \rho_t^{\otimes p-1}\otimes \mathbb{I} \cdot (I - \eta M_D) \cdot \rho_t^{\otimes p} = \rho_t^{\otimes p} - \eta \rho_t^{\otimes p-1}\otimes \mathbb{I} \cdot M_D  \cdot\rho_t^{\otimes p}.
\end{align}
We recall that the original formulation of $M_D$:
$$  M_D = \sum_{\alpha=1}^K\sum_{j=1}^p \Big( \bigotimes_{i=1,i\neq j}^p A_{i}^{\alpha} \Big) \bigotimes A_j^{\alpha}. $$
Therefore, we have 
\begin{align}
    \rho_t^{\otimes p-1}\otimes \mathbb{I}\cdot M_D \cdot\rho_t^{\otimes p} &= \sum_{\alpha=1}^K\sum_{j=1}^p  \Big(  \bigotimes_{i=1,i\neq j}^p \ket{x_t}\bra{x_t} A_{i}^{\alpha} \ket{x_t} \bra{x_t} \Big) \bigotimes A_j^{\alpha} \ket{x_t}\bra{x_t} \\
    &= \sum_{\alpha=1}^K\sum_{j=1}^p  \Big(  \bigotimes_{i=1,i\neq j}^p \bra{x_t} A_{i}^{\alpha} \ket{x_t} \ket{x_t} \bra{x_t} \Big) \bigotimes A_j^{\alpha} \ket{x_t}\bra{x_t} \\
    &= \ket{x_t}\bra{x_t}^{\otimes p-1} \otimes D(\ket{x_t}) \ket{x_t}\bra{x_t}.
\end{align}
We thus arrive at the following action,
\begin{align}
  P_t \ket{x_t}^{\otimes p} &=   \Big(\frac{\pi}{4 }\Big)^{2p-1} \frac{1}{sp} \ket{x_t}^{\otimes p-1 } \otimes \Big(I_n- \eta D(\ket{x_t})\Big) \ket{x_t},
\end{align}
where $I_n$ denotes specifically the identity matrix of dimension $n \times n$, which is the same dimension as that of $\ket{x_t}$. As we know that $U_t$ can be explicitly written out as:
\begin{align}
    U_t = \ket{\bf 0}\bra{\bf 0} \otimes P_t + \cdots, 
\end{align}
its  action  on an arbitrary state $\ket{\bf 0}\ket{\phi}$ (the dimension of $\ket{\phi}$ is $n^p$) is:
\begin{align}
    U_t \ket{\bf 0}\ket{\phi} = \ket{\bf 0} P_t \ket{\phi} + \sum_{j \neq \textbf{0}} \ket{j} \ket{\rm Garbage_j}.
    \label{eqn: 57}
\end{align}
If we choose $\ket{\phi} = \ket{x_t}^{\otimes p}$, and perform measurement on the first register, post-selecting the outcome on seeing $\ket{\bf 0}$, then we obtain: 
$$ \frac{1}{C} \ket{x_t}^{\otimes p-1} \otimes (I_n-\eta D(\ket{x_t})) \ket{x_t}, $$
which contains our desired state in the last register. In the above, $C$ is a normalization factor, which is: 
\begin{align}
    C^2 &= || (I_n- \eta D(\ket{x_t}))\ket{x_t} ||^2  \\
    &= (1 - 2 \eta \bra{x_t}D (\ket{x_t}) \ket{x_t} + \eta^2 \bra{x_t} D^2(\ket{x_t}) \ket{x_t} ),
\end{align}
and it is proportional to the probability of successfully obtaining the outcome $\ket{\bf 0}$ on the first register. As also proved in~\cite{rebentrost2019quantum}, the factor $C$ is lower bounded by:
\begin{align}
    C^2 \geq (1 - \eta \sqrt{ \bra{x_t} D^2(\ket{x_t} )\ket{x_t} }  )^2.
\end{align}

Previously, we have made the assumption that the norm of $D$ satisfies that $|D| \leq p$, which means that 
$$ \sqrt{ \bra{x_t} D^2(\ket{x_t}) \ket{x_t} } \leq \max_{y} ||D \ket{y}||  \leq p. $$
If we choose $\eta$ to satisfy $ 0 < \eta < 1/(2p)$, then we have:
$C^2 \geq {1}/{4}$. 
As in the previous version, we can use the quantum amplitude amplification ~\cite{brassard2002quantum} to improve the success probability \revise{quadratically better}. Therefore, altogether, it yields the following lower bound:
$$ p_{\rm success} \geq \Big(\frac{\pi}{ 4 }\Big)^{2p-1} \frac{1}{4sp},   $$
and the total number of repetitions required is 
$$\mathcal{O}\Big( sp \Big(\frac{ 4}{\pi}\Big )^{2p-1} \Big), $$
where we have absorbed a factor $\pi/2$ into the $\mathcal{O}$ scaling. \revise{Remind that the $\epsilon$-approximation block encoding of $M_D/( sp )$ is  obtained in time complexity $\mathcal{O} \big(p^2\log(n) + p \log^{2.5}\frac{1}{\epsilon} \big)$}. The tolerance in all steps is set to be $\epsilon$ for convenience. We summarize the single-gradient step in the following theorem. 
\begin{theorem}
\label{thm: secondversion}
   Given $ \mathcal{O}( \revise{\frac{p}{\epsilon} \log\frac{1}{\epsilon}} ) $ copies of $\ket{x_t}$, there exists a quantum algorithm that produces an $\epsilon$-tolerance of the updated state that corresponds to 
   $$ x_{t+1} = \ket{x_t} - \eta D(\ket{x_t}) \ket{x_t}. $$
   The running time for obtaining the state $\ket{x_{t+1}}$ from copies of $\ket{x_t}$ is 
$$ \mathcal{O} \Big( \big( p^3\log(n) + \frac{p^2}{\epsilon} \log(\frac{1}{\epsilon})  \revise{+  p \log^{2.5}\frac{1}{\epsilon} } \big) s \big(\frac{4 }{\pi} \big)^{2p-1}  \Big).  $$
\end{theorem}

In~\cite{rebentrost2019quantum} \revise{(according to Result 5)}, a single gradient step requires $ \mathcal{O}\Big( \revise{\frac{p^5}{\epsilon^4}} \Big ) $ copies of $\ket{x_t}$, with a further running time $ \mathcal{O}\Big( \revise{\frac{p^5}{\epsilon^4} s \log(n) }\Big)$ in order to obtain the updated state $\ket{x_{t+1}}$.  One can see that in our approach there is a polynomial improvement (i.e., saving) in the number of copies required. For the total running time, our method achieves significant improvement over the dependence on the error tolerance $\epsilon$, but with the same $\mathcal{O}(\log n)$ dependence. Regarding the factor $p$, we observe that our method has a factor $(\pi/2)^{2p-1}$ grows exponentially \revise{with respect to} $p$. To assess the whether this is a disadvantage, we must compare the growth of the two functions: $p^5$ and $p^3 (4/\pi)^{2p-1}$ and the regime of $p$. For example, from numerical evaluation (see Fig.~\ref{fig: compare} below), we observe that for $1 < p< 10$, the value of $p^3 (4/\pi)^{2p-1}$ is smaller than $p^5$. Therefore, this is the regime where our method is overall better than that of~\cite{rebentrost2019quantum}. 
\begin{figure}[H]
    \centering
    \begin{tikzpicture}[scale = 1.0]
        \begin{axis}[
            xlabel = {$p$},
            ylabel = {$y$},
            axis x line=bottom,
                axis y line=left
        ]
            \addplot[domain=0:12, thick, blue] {x*x*x*x*x};
            \addlegendentry{$y=p^5$}

            \addplot[domain=0:12, thick, red] {x*x*x * (4/pi)^(2*x-1)};
            \addlegendentry{$y =p^3 \Big(4/\pi\Big)^{2p-1} $}

            \addplot[purple, thick, dashed] coordinates {(10, 0) (10, 2*10^5)};
        %\addlegendentry{$x = 10$}
        \end{axis}
    \end{tikzpicture}
    \caption{Plot of function $p^5$ and $p^3 \Big( \frac{4}{\pi} \Big)^{2p-1}$  for the scenario that $C_1$ is assumed to be equal to $C_2$. However, below, we show that $C_1 \gg C_2$, which means that in reality the range of $p$ such that $C_1 p^5  \geq C_2 p^3 \Big( \frac{4}{\pi} \Big)^{2p-1} $ can be larger. }
    \label{fig: compare}
\end{figure}
\revise{Here, we discuss an important subtlety. The comparison above between $p^5$ and $p^3 (4/\pi)^{2p-1}$ was ``raw'', as we used only the big-$\mathcal{O}$ notation and did not take into account the detailed scaling. The more precise scaling of Ref.~\cite{rebentrost2019quantum} including the coefficients can be $C_1 p^5$, and our scaling is $C_2 p^3 (4/\pi)^{2p-1}$. Although $ p^5 > p^3 (4/\pi)^{2p-1}$ for $1<p<10$, as illustrated in Fig.~\ref{fig: compare}, it is not guaranteed to be the same after considering these coefficients. 
We provide a more detailed discussion showing that the constant factor within the big-$\mathcal{O}$ of Theorem~\ref{thm: secondversion} is significantly smaller than that of~\cite{rebentrost2019quantum}, i.e., $C_2 \ll C_1$. We recall the first two steps of Algorithm~\ref{method: algorithm3}, where we need to take multiple copies of $\ket{x_t}$, to produce an approximation of operator $\exp(-i \ket{x_t}\bra{x_t}) $ and obtain an approximated block encoding of $\pi \ket{x_t}\bra{x_t}/4$. According to~\cite{rebentrost2014quantum}, the number of steps, or time complexity, required to implement $\exp(-i\ket{x_t}\bra{x_t} t)$ acting on $\ket{\bf 0}$ to accuracy $\epsilon$ is 
\begin{align}
    N = \frac{ \Big| \ket{x_t}\bra{x_t}- \ket{\bf 0}\bra{\bf 0} \Big|^2 t^2 }{ \epsilon},
\end{align}
which is also the sample number of $\ket{x_t}$ required. As $|.|$ refers to the operator norm, it is obvious that $\Big| \ket{x_t}\bra{x_t}- \ket{0}\bra{0} \Big|^2 < 1$. We also choose $t=1$, so $N = \mathscr{C}_1 \frac{1}{\epsilon}$, where $\mathscr{C}_1 < 1$. The next step is to use $\exp(-i \ket{x_t}\bra{x_t})$ combined with Lemma~\ref{lemma: logU} to construct the block encoding of $\frac{\pi}{4}\ket{x_t}\bra{x_t}$. Based on~\cite{gilyen2019quantum}, the underlying mechanism of Lemma~\ref{lemma: logU} is Lemma~\ref{lemma: theorem56}, i.e., choosing a polynomial that approximates the function $\frac{2}{\pi}\arcsin(x)$ on the interval $(-1/2,1/2)$. According to Lemma 70 in~\cite{gilyen2019quantum}, the polynomial that approximates the desired function has degree $2 \log \frac{1}{\epsilon}$. Thus, Lemma~\ref{lemma: logU} uses $2 \log \frac{1}{\epsilon} $ (controlled) unitary $\exp(-i\ket{x_t}\bra{x_t})$ and its inverse. As mentioned above, each unitary $\exp(-i\ket{x_t}\bra{x_t})$ can be implemented in $N = \mathscr{C}_1 \frac{1}{\epsilon}$ time complexity. So in total it takes $4 \mathscr{C}_1 \frac{1}{\epsilon}\log\frac{1}{\epsilon}  $ time complexity, and also the same amount of sample complexity for obtaining the block encoding of $ \frac{\pi}{4}\ket{x_t}\bra{x_t}$. The next step is to use Lemma~\ref{lemma: As} to construct the block encoding of $\frac{A}{s}$, and then use Lemma~\ref{lemma: sumencoding} to construct the block encoding of $M_D = \sum_{j=1}^p \frac{1}{ps} Q_j A Q_j$, where each $Q_j$ is a SWAP operation between two $\log(n)$-qubit registers. The time complexity of obtaining a block encoding of $A/s$ is $\mathcal{O}\big(  \log (n^p) + \log^{2.5} \frac{1}{\epsilon}\big)$. We use $C_2$ to denote the prefactor in the big-$O$ notation, i.e., the time complexity of producing $A/s$ is then $C_2 \big( \log(n^p) + \log^{2.5} \frac{1}{\epsilon}\big)$. Lemma~\ref{lemma: sumencoding} uses $p$ block encodings of $A/s$ to construct $M_D$, so the total complexity is $ p C_2 \big( \log(n^p) + \log^{2.5} \frac{1}{\epsilon}\big)$. Eventually we construct the unitary block encoding of $\big(\frac{\pi\rho_t}{4}\big)^{\otimes p-1}\otimes \mathbb{I} \cdot \frac{(I - \eta M_D) }{sp} \cdot (\frac{ \pi \rho_t}{4 } )^{\otimes p}$, before apply such unitary to the state $\ket{\bf 0}\ket{x_t}^{\otimes p}$, perform measurement and post-select $\ket{\bf  0}$, resulting in a total complexity 
\begin{align}
    ps \big( \frac{4}{\pi}\big)^{2p-1} \Big( p C_2 \big( \log(n^p) + \log^{2.5} \frac{1}{\epsilon}\big) + 4\mathscr{C}_1 \frac{1}{\epsilon}\log \frac{1}{\epsilon} \Big).
\end{align}
We recall from the above that $\mathscr{C}_1$ is less than 1, the above complexity is upper bounded by 
\begin{align}
    s \big( \frac{4}{\pi}\big)^{2p-1}  \Big(C_2 \big( p^3\log(n) + p^2\log^{2.5} \frac{1}{\epsilon}\big) + 4 p\frac{1}{\epsilon}\log \frac{1}{\epsilon} \Big).
\end{align}
Next, we consider the method in~\cite{rebentrost2019quantum}. While their algorithm is more complicated, we point out that there is a step where they need to obtain $\exp(-iM_D t)$. They achieve such a goal by approximating the  operator as
\begin{align}
    \exp(-i M_D t) \approx \prod_{j=1}^p Q_j \exp(-iA t) Q_j.
\end{align}
While  Ref.~\cite{rebentrost2019quantum} used a more expensive simulation technique, the optimal cost for approximating $\exp(-iA t)$ has been established in~\cite{low2017optimal,low2019hamiltonian, gilyen2019quantum}, which uses the block encoding of $\frac{A}{s}$ multiple times. It means that at the very least, the complexity of~\cite{rebentrost2019quantum} is scaling as $ \alpha C_2 \big( \log(n^p) \big)$, where $\alpha$ roughly counts the times the blocking encoding of $A/s$ is used and it should be greater than $1$. Hence, the prefactor $C_1$ in the complexity $\mathcal{O}\Big( p^5 \frac{s}{\epsilon^4} \log n \Big) \equiv C_1 p^5 \frac{s}{\epsilon^4} \log n $  is greater than $C_2$, which implies that $C_1 p^5 > C_2 p^3 (\frac{4}{\pi})^{2p-1}$ at least in the interval $1<p<10$. 
}

In the above, we discuss a single-step gradient descent. Multiple-step gradient descent is a purely straightforward execution from the single-step gradient descent. Therefore, the total cost of our improved method is still exponential in the number of iteration steps $T$, similar to~\cite{rebentrost2019quantum}. However, the enhancement of our method for the whole multi-step procedure is substantial, as it is already more efficient in each iteration step, as discussed above.

\section{Discussion And Conclusion}\label{sec:discussconclude}
Here, we discuss some possible extensions that are directly applicable from our methodology. 

\subsection{From Largest Eigenvalue to Multiple Eigenvalues}
Interestingly, the method introduced in this work via the block encoding is relatively simple compared to previous constructions, but yields a substantial speedup. The first problem we have discussed is finding the largest eigenvalue based on the quantum power method, as has been previously discussed by us in~\cite{nghiem2022quantum}, and its inverse version (by replacing the original matrix $A$ by its inverse $A^{-1}$) can be used to find the minimum eigenvalue (in magnitude). It is well-known that the minimum eigenvalue problem is a very important problem, for instance, the ground-state energy of a given system. Therefore, improving this work suggests a more efficient way to achieve the other task better by reducing the cost induced by the iteration steps. While in~\cite{nghiem2022quantum},  a hybrid Krylov subspace method was introduced to find multiple eigenvalues, here, we point out another simpler and potentially more efficient way to find multiple eigenvalues when the given matrix $A$ is positive-definite, i.e., having positive eigenvalues.  

Denote the eigenvalues of $A$ (in increasing magnitude) $0 < \lambda_1 < \lambda_2, ..., \lambda_n < 1$ (note $n$ is the dimension of $A$ and of course we consider $A$ to be Hermitian). Suppose we already find the largest eigenvalue $\lambda_n$ of $A$, e.g., via the improved quantum power method. If we compute the difference $\lambda_n - \lambda_i$ for $i = 1,2,..., n-1$, then $\lambda_n - \lambda_1$ has the largest magnitude. It means that if we consider the matrix: 
$$ \lambda_{n} I_{n \times n} - A. $$
Then this matrix has $\lambda_n - \lambda_1$ to be its largest eigenvalue.    However, this matrix has a 0 eigenvalue (since $\lambda_n - \lambda_n = 0$), which might cause some issues. For instance, if one randomizes an initial vector, say $x_0$, that happens to be the eigenvector with the zero eigenvalue, then we cannot execute the method, as $Ax_0 = 0$. A simple way to avoid  this issue is to slightly shift the spectrum by a constant $\Delta$, i.e., we consider instead the matrix 
$$ (\lambda_n + \Delta) I_{n\times n} -A.  $$
We note that we still need to choose $\Delta$ such that $\lambda_n + \Delta \leq 1$. Then again, this matrix has $\lambda_n + \Delta - \lambda_1$ to be its largest eigenvalue. Since we know $\lambda_n$ (and apparently $\Delta$), we can estimate $\lambda_1$, which is the minimum eigenvalue. Therefore, we can employ the improved quantum power method here to find $\lambda_n + \Delta -\lambda_1$. As we mentioned previously, the block encoding of $A/s$ is easily constructed, and hence the block encoding of $(\lambda_{n} + \Delta) I_{n \times n}/s $ is also easy to construct. For example, one considers the matrix $R_Y$ 
\begin{align}
   R_Y(\theta) = \begin{pmatrix}
        \cos(\theta/2) & -\sin(\theta/2) \\
        \sin(\theta/2) & \cos(\theta/2) 
    \end{pmatrix}.
\end{align}
If we choose $\cos (\theta/2) = (\Delta +\lambda_n)/s$ and construct its tensor product with, for instance, $I_{n \times n}$, then we obtain a unitary block encoding of a $n\times n$ matrix that contains only $\lambda_n/s$ on the diagonal. The construction of the unitary block encoding of 
$$ \frac{\lambda_n+\Delta}{s} I_{n \times n} - \frac{A}{s} = \frac{1}{s}\big((\lambda_n+\Delta ) I_{n\times n} -A \big) $$
is doable due to Lemma \ref{lemma: sumencoding}.
 
Note that as $\lambda_n-\lambda_1$ is the largest eigenvalue of the matrix $\lambda_n I_{n\times n}-A$, it is straightforward to see that $\lambda_n - \lambda_{n-1}$ is its minimum (non-zero) eigenvalue. Therefore, one can repeat the same process with the new matrix above to find $\lambda_n - \lambda_{n-1}$, which gives us the value of $\lambda_{n-1}$ -- the second largest eigenvalue of $A$. Hence, doing this multiple times allows us to find multiple eigenvalues of $A$.

\subsection{Newton Iteration Method for Matrix Inversion}
We have provided two perspectives on the quantum gradient descent method.   While the second version is clearly a direct improvement on what has been done previously in~\cite{rebentrost2019quantum}, the first one is more or less an indirect procedure inspired purely by block encodings.  Here, we wish to point out that the idea presented in the first version actually has a very relevant application: finding the inverse of some matrix $A$. We recall that the main spirit of this approach is to encode our matrices into  larger ones, and by using simple operations of the block-encoded matrices (see Lemmas~\ref{lemma: product}~\&~\ref{lemma: sumencoding}), a simple iterative procedure is carried out. It turns out that there is a similar method, i.e., the Newton iteration method for finding the inverse of a given matrix, to which we can also apply our idea straightforwardly. This is a well-known classical approach, and its proof of performance guarantee can be found in any standard literature~\cite{kelley1995iterative}. We simply summarize the procedure first.

Suppose we are given matrix $A$ and hope to find its inverse $A^{-1}$. We begin with a random guess $X_0 = \alpha A^T$ where $\alpha$ is some small parameter. At the $t$-th iteration step, the matrix is updated as follows:
\begin{align}
    X_{t+1} = 2 X_t - X_t A X_t. 
    \label{eqn: 61}
\end{align}

In order to construct a quantum approach, we apply the idea from the first version, described in Sec.~\ref{sec: firstversion}. First, we note that the black box access to entries of $A$ allows us to construct the unitary block encoding of $A/s$, where $s$ is the sparsity, which directly yields the block encoding of $A^T/s$. To insert the factor $\alpha$, we choose $\cos(\theta/2) = \alpha$  in the RY rotational gate 
\begin{align}
   R_Y(\theta) = \begin{pmatrix}
        \cos(\theta/2) & -\sin(\theta/2) \\
        \sin(\theta/2) & \cos(\theta/2) 
    \end{pmatrix}
\end{align}
 and construct the tensor product $R_Y(\theta) \otimes A^T$, which contains the matrix $\alpha A^T$ in the top left corner, then we obtain exactly the unitary block encoding of $\alpha A^T \equiv X_0$. The remaining task is straightforward as we simply use Lemmas~\ref{lemma: sumencoding}~\&~\ref{lemma: product} to iteratively update the matrix according to Eqn.~(\ref{eqn: 61}). Hence, we obtain the desired approximation of the inverse of $A$ encoded in a bigger unitary, for which subsequent operations can be executed depending on the application context. 

\subsection{Conclusion } 
We have successfully introduced more efficient methods for two different problems: estimating the largest eigenvalue and performing the quantum gradient descent, \revise{which are two very useful computational problems enjoying a wide range of applications.} The underlying framework of our algorithm is the powerful unitary block encoding~\cite{gilyen2019quantum, low2017optimal, low2019hamiltonian}, which has recently transformed quantum algorithms in a very fundamental way, as it provides a unified viewpoint for all previously proposed algorithms. \revise{Such a versatile framework enables us to avoid the necessity of repeating measurements in order to obtain the desired states. Instead, we can directly manipulate and operate on the pre-measured, or intermediate state, by block-encoding them into a unitary transformation. The key information, e.g., the largest eigenvalue, can be revealed by performing a fixed amount of measurement on another ancillary system instead, thus avoiding the exponential cost imposed in~\cite{nghiem2022quantum}. For the second problem, which is the gradient descent, we have seen that the block encoding technique allows us to side-step many sophisticated and technical step in~\cite{rebentrost2019quantum}. Instead of relying on measurement and post-selected state, we can directly block-encode the density state and form the gradient operator on the block-encoding subspace, resulting in a much neater yet more efficient framework for performing the gradient descent. } Our work thus contributes as another instance of demonstrating the efficacy of the unitary block encoding method, following the success in~\cite{low2017optimal}, \cite{rall2020quantum},~\cite{chakraborty2018power}, and~\cite{gilyen2022quantum}. \revise{In fact, we have not even used any sophisticated techniques of the quantum singular value transformation.} \revise{Rather,} it is remarkable that our method utilizes only the block-encoded matrices and elementary operations between them, \revise{such as performing scalar multiplication and building linear combination} to construct an efficient procedure that removes some exponential scaling from the original versions. This highlights the subtle but extremely useful ideas from quantum signal processing, or quantum singular value transformation, that allow the handling of arbitrary matrices, which is not necessarily unitary.  It is thus of interest to explore further the power of such block encoding or quantum singular value transformation framework, and we hope our work to motivate further such exploration into various contexts.

\begin{acknowledgements}
\revise{We thank an anonymous reviewer for insightful suggestions.} This work was supported in part by the US Department of Energy, Office of Science, National Quantum Information Science Research Centers, Co-design Center for Quantum Advantage (C2QA) under contract number DE-SC0012704. We also acknowledge the support by a Seed Grant from Stony Brook University’s Office of the Vice President for Research and by the Center for Distributed Quantum Processing. 
\end{acknowledgements}

\bibliography{ref.bib}{}
\bibliographystyle{unsrt}

\clearpage
\newpage
\onecolumngrid

\appendix
\section{Preliminaries}
\label{sec: preliminaries}
Here, we summarize the key recipes used in our subsequently described quantum algorithms. We keep the statements here concise and precise, with their proofs/constructions referred to the original works or the appendices of this paper.  Readers familiar with these results can skip this section. Throughout the work, we use $||\cdot||$ to denote the matrix/vector norm, $|.|$ to denote the absolute value of numbers. 

\begin{definition}[Block Encoding Unitary]~\cite{low2017optimal, low2019hamiltonian, gilyen2019quantum}
\label{def: blockencode} 
Let $A$ be some Hermitian matrix of size $N \times N$ whose matrix norm $|A| < 1$. Let a unitary $U$ have the following form:
\begin{align*}
    U = \begin{pmatrix}
       A & \cdot \\
       \cdot & \cdot \\
    \end{pmatrix}.
\end{align*}
Then $U$ is said to be an exact block encoding of matrix $A$. Equivalently, we can write:
\begin{align*}
    U = \ket{ \bf{0}}\bra{ \bf{0}} \otimes A + \cdots
\end{align*}
where $\ket{\bf 0}$ simply denotes the extra ancilla system. In the case where the $U$ has the form 
$$ U  =  \ket{ \bf{0}}\bra{ \bf{0}} \otimes \Tilde{A} + \cdots, $$
where $|| \Tilde{A} - A || \leq \epsilon$, then $U$ is said to be an $\epsilon$-approximated block encoding of $A$.
\end{definition}
The above definition has multiple natural corollaries. First, an arbitrary unitary $U$ block encodes itself. Suppose $A$ is block encoded by some matrix $U$, then $A$ can be block encoded in a larger matrix by simply adding ancillas (which have dimension $m$). Note that $\Ibb_m \otimes U$ contains $A$ in the top-left corner, which is a block encoding of $A$ again by definition. Further, it is almost trivial to block encode the identity matrix of any dimension. For instance, we consider $\sigma_z \otimes \Ibb_m$ (for any $m$), which contains $\Ibb_m$ in the top-left corner. 

From the above definition, suppose $\ket{\phi}$ is arbitrary state having the same dimension as $A$, we notice that:
\revise{
\begin{align}
    \label{eqn: action}
    U \ket{\bf 0}\ket{\phi}  &= \ket{\bf 0} A\ket{\phi} + \sum_{ \textbf{j} \neq \textbf{0}} \ket{\bf j} \ket{\rm Garbage}_j \\
\end{align}
where $\ket{\rm Garbage}_j $ is an improperly normalized, irrelevant state. 
}

The proof of the above lemma is given in~\cite{gilyen2019quantum} (see their Lemma 45). \\

\begin{lemma}[Block Encoding of Product of Two Matrices]
\label{lemma: product}
    Given the unitary block encoding of two matrices $A_1$ and $A_2$, an efficient procedure exists that constructs a unitary block encoding of $A_1 A_2$.
\end{lemma}

The proof of the above lemma is given in appendix~\ref{sec: product}. \\

\begin{lemma}[\cite{camps2020approximate}]
\label{lemma: tensorproduct}
    Given the unitary block encoding $\{U_i\}_{i=1}^m$ of multiple operators $\{M_i\}_{i=1}^m$ (assumed to be exact encoding), then there is a procedure that produces the unitary block encoding operator of $\bigotimes_{i=1}^m M_i$, which requires a single use of each $\{U_i\}_{i=1}^m$ and $\mathcal{O}(1)$ SWAP gates. 
\end{lemma}

The above lemma is a result from~\cite{camps2020approximate}. 
\begin{lemma}
\label{lemma: As}
    Given oracle access to an $s$-sparse matrix $A$ of dimension $n\times n$, then an $\epsilon$-approximated unitary block encoding of $A/s$ could be prepared with gate/time complexity $\mathcal{O}\big(\log n + \log^{2.5}(\frac{1}{\epsilon})\big)$.
\end{lemma}

This is a standard result from previous works~\cite{low2017optimal, gilyen2019quantum, low2019hamiltonian}. In fact, the optimal Hamiltonian simulation algorithm outlined in~\cite{low2019hamiltonian,low2017optimal} essentially makes use of the oracle access to entries of a given Hamiltonian $H$ to construct the unitary block encoding of $H$, followed by a series of transformations on such blocks to approximate the desired operator, i.e., $\exp(-i H t)$. 

\revise{\begin{lemma}[Scaling Block encoding] 
\label{lemma: scale}
    Given a block encoding of some matrix $A$ (as in~\ref{def: blockencode}), then the block encoding of $A/p$ where $p > 1$ can be prepared with an extra $\mathcal{O}(1)$ cost.  
\end{lemma}}
%\revise{
%\begin{lemma} 
%\label{lemma: qsvt}
%[\cite{gilyen2019quantum} Theorem 56]
%\label{lemma: theorem56}  (Polynomial Transformation)
%Suppose that $U$ is an
%$(\alpha, a, \epsilon)$-encoding of a Hermitian matrix $A$. (See Definition 43 of~\cite{gilyen2019quantum} for the definition.)
%If $P \in \mathbb{R}[x]$ is a degree-$d$ polynomial satisfying that
%\begin{itemize}
%\item for all $x \in [-1,1]$: $|P(x)| \leq \frac{1}{2}$,
%\end{itemize}
%then, there is a quantum circuit $\tilde{U}$, which is an $(1,a+2,4d \sqrt{\frac{\epsilon}{\alpha}})$-encoding of $P(A/\alpha)$ and
%consists of $d$ applications of $U$ and $U^\dagger$ gates, a single application of %controlled-$U$ and $\mathcal{O}((a+1)d)$
%other one- and two-qubit gates.
%\end{lemma}}

\section{Product of Block Encoding}
\label{sec: product}
Here we explicitly show that, given unitary block encoding of matrices $A_1$ and $A_2$ respectively, one can construct the unitary block encoding of $A_1 A_2$, or $A_2 A_1$ with a few modest steps. While it has been established in the original work~\cite{gilyen2019quantum}, we present a proof here for completeness. \\

Let $U_1, U_2$ be block encoding of $A_1, A_2$, i.e, we have:
\begin{align*}
    U_{1,2} = \begin{pmatrix}
       A_{1,2} & \cdot \\
       \cdot & \cdot \\
    \end{pmatrix}.
\end{align*}
Equivalently, we can write $U$ (we drop the subscript for now, as it is not too important): 
\begin{align*}
    U = \ket{ \bf{0}}\bra{ \bf{0}} \otimes A + \cdots
\end{align*}
We observe the following property
\begin{align}
    U \ket{\bf 0} \ket{\phi} = \ket{\bf 0} A\ket{\phi} + \sum_{j \neq \textbf{0}} \ket{j} \ket{\phi_j},
\end{align}
where $\ket{\phi}$ and all $\ket{\phi_j}$'s (essentially garbage states) share the same dimension as matrix $A$. For a reason that will be clear later on, we borrow an extra qubit initialized in $\ket{0}$ and rewrite the above equation as:
\begin{align}
   \mathbb{I} \otimes U \ket{1}\ket{\bf 0} \ket{\phi} = \ket{0} \ket{\bf 0} A\ket{\phi} + \ket{0} \sum_{j \neq \textbf{0}} \ket{j} \ket{\phi_j}.
\end{align}

We use $X$ gate to flip the ancilla qubit (to $\ket{1}$) to obtain the state: 
\begin{align*}
    \ket{1} \ket{\bf 0} A\ket{\phi} + \ket{1} \sum_{j \neq \textbf{0}} \ket{j} \ket{\phi_j}.
\end{align*}
Denote the whole above unitary process, which maps $\ket{0}\ket{\bf 0} \ket{\phi}$ to $\ket{1} \ket{\bf 0} A\ket{\phi} + \ket{1} \sum_{j \neq \textbf{0}} \ket{j} \ket{\phi_j} $ as $\mathcal{U}$. For matrix $A_1$, given a computational basis state $\ket{i}$, we have the corresponding action $\mathcal{U}_1$:
\begin{align}
    \mathcal{U}_1  \ket{\bf 0} \ket{i} =  \ket{1} \ket{\bf 0} A_1\ket{i} + \ket{1} \sum_{j \neq \textbf{0}} \ket{j} \ket{\phi_j}.
\end{align}
For matrix $A_2$, we will use the conjugate transpose of the original $U_2$ (without the extra CNOT step that flips the ancilla) to obtain:
\begin{align}
   U_2^\dagger  \ket{\bf 0} \ket{k} =  \ket{0} \ket{\bf 0} A_2^\dagger \ket{k} + \ket{0} \sum_{j \neq \textbf{0}} \ket{j} \ket{\phi_j}.
\end{align}

Now, for the above state, we use the register $\ket{\bf 0}$ as controlling qubits, aiming to flip the first qubit $\ket{0}$ to $\ket{1}$. In other words, from the above state, we obtain:
\begin{align}
    \ket{1} \ket{\bf 0} A_2^\dagger \ket{k} + \ket{0} \sum_{j \neq \textbf{0}} \ket{j} \ket{\phi_j}.
\end{align}

We denote the process containing $U_2$ and the controlled-X step above as $\mathcal{U}_2$. To summarize, we have
\begin{align}
    \mathcal{U}_2 \ket{0}\ket{\bf 0}\ket{k} =  \ket{1} \ket{\bf 0} A_2^\dagger \ket{k} + \ket{0} \sum_{j \neq \textbf{0}} \ket{j} \ket{\phi_j}.
\end{align}
It is straightforward to observe that, thanks to the orthogonality of computational basis states, the inner product: 
\begin{align}
    \bra{0}\bra{\textbf{0}, k} \mathcal{U}_2^\dagger \mathcal{U}_1 \ket{0}\ket{ \textbf{0}, i } = \bra{k} A_2 A_1 \ket{i} \equiv (A_2 A_1)_{ki},
\end{align}
which is exactly the definition of unitary block encoding. Therefore, we have successfully constructed the block encoding of $A_2 A_1$. The procedure is the same for the reverse order $A_1 A_2$, except that we need to reverse the role of $\mathcal{U}_{1,2}$ from the above procedure.  

\section{Norms of Solution}
In this section, we show that if the initial solution $x_0$ has a bounded norm, then the norm of the solution at any time step $t$ is guaranteed to be less than unity. Our proof is borrowed from the analysis of success probability in~\cite{rebentrost2019quantum}. Once we begin with $x_0$, we have that:
\begin{align}
    x_1 = x_0 - \eta D(x_0) x_0.
\end{align}
So the norm is:
\begin{align}
    ||x_1||^2 &= ||x_0||^2 - 2 \eta x_0^T \cdot D(x_0)x_0  + \eta^2 || D(x_0) x_0 ||^2 \\
    & \leq ||x_0||^2 + 2 \eta ||x_0|| ||D(x_0)x_0 || + \eta^2 || D(x_0) x_0 ||^2 \\
    & \leq ||x_0||^2 ( 1 + \eta ||D|| )^2 \\
    & \leq 4 ||x_0||^2,
\end{align}
where the last line comes from the fact that one can choose $\eta$ to be smaller than $1/||D|| < 1/p$. Iteratively continuing the procedure, we have that at the $t$-th iteration:
\begin{align}
    ||x_t ||^2 \leq  4^t ||x_0||^2.
\end{align}
If we choose $x_0$ such that:
%\begin{align}
    $||x_0||^2 \leq 1/4^t$,
%\end{align}
then, the norm of $x_t$ is guaranteed to be less than unity, as we desired. The question now is how to generate $x_0$ satisfying such a condition. A simple solution is that we choose to generate a vector $X$ of dimension $M > n$, and pick the first $n$ amplitudes so that their squared sum is less than $1/4^t$. Such a state could be prepared with a method outlined in~\cite{prakash2014quantum}. Another simple way to prepare such a state apparently exists. If we use $2t$ qubits initialized in $\ket{0}$ and apply Hadamard gates, we obtain the state:
\begin{align*}
    \ket{\phi} = \frac{1}{\sqrt{2^{2t}}} \sum_{i}^{4^t} \ket{i}.
\end{align*}
If we choose $t$ such that $2t > \log(n)$, then the above summation has $4^t > n$ terms. If we consider only the first $n$ amplitudes, then their squared sum is $n/4^t$. We simply need to choose $t' = t + \log(n)/2$, then the norm is $1/4^t$, which is exactly the desired norm. Given that we can prepare such $M$-dimensional state, Lemma~\ref{lemma: improveddme} allows us to construct the block encoding of $X X^T$ of dimension $M\times M$. If we limit ourselves to the top left block of dimension $n \times n$, then this is the operator $x_0 x_0^T$ that we desired.

\section{Proof of Lemma \ref{lemma: 11}}
\label{sec: proof11}
We remind the lemma we wish to prove:

\noindent {\bf Lemma 11} \textit{ Given block encoding of $(x x^T)^{\otimes p-1} \otimes \frac{D(x)}{ps}$, then it is possible to obtain the block encoding of $D(x)/ps$ in $\mathcal{O}( \gamma^{2(p-1)} ps )$ where $\gamma$ is some  constant. }

We first define several symbols. Let $P$ denote the block encoding of $X \equiv (x x^T)^{\otimes p-1} \otimes \frac{D(x)}{ps}$. Let $A$, $B$, and $C$ denote the registers corresponding to the auxiliary system of block encoding (see the definition~\ref{def: blockencode}), $(xx^T)^{\otimes p-1}$, and $D(x)/ps$, respectively. Let $U$ denote some unitary of dimension $n \times n$. Then, we have the following:
\begin{align}
    P (I_A \otimes U^{\otimes p-1}\otimes I ) \ket{\bf 0}_A \ket{\bf 0}_B \ket{k}_C = \ket{\bf 0}_A X (U^{\otimes p-1})\ket{\bf 0}_B \ket{k}_C + \ket{\Phi_{\perp}},
\end{align}
where $\ket{\Phi_{\perp}}$ satisfies that $ \ket{\bf 0}_A \bra{\bf 0}_A \otimes I \cdot \ket{\Phi_{\perp}} = 0$. Let us note the following:
\begin{align*}
    U^{\otimes p-1} \ket{\bf 0}_B = U^{\otimes p-1} \ket{0}^{\otimes p-1} = \ket{\alpha}^{\otimes p-1},
\end{align*}
and define the symbol $\beta = \bra{\alpha} xx^T \ket{\alpha}$. It is then straightforward to see that:
\begin{align}
  \bra{\bf 0}_B \bra{i}_C  U^{\dagger^{\otimes p-1}} X U^{\otimes p-1} \ket{\bf 0}_B \ket{k}_C =  \beta^{p-1} \bra{i}_C \frac{D(x)}{ps} \ket{k}_C = \beta^{p-1} (D(x)/ps)_{ik},
\end{align}
where the last term comes from the definition of a matrix. Therefore, the unitary $(I_A \otimes U^{\dagger^{\otimes p-1}}\otimes I ) P (I_A \otimes U^{\otimes p-1} \otimes I ) $ is the unitary block encoding of $\beta^{p-1} D(x)/ps$. Thanks to the preamplification technique introduced in~\cite{gilyen2019quantum}, one can use the block encoding above roughly $\mathcal{O}((1/\beta)^{p-1})$ times to remove such a factor. Therefore, the factor $\gamma$ in the stated lemma is $1/\sqrt{\beta}$. We remark that the factor $\beta = \bra{\alpha} x x^T \ket{\alpha}$ can be estimated using the Hadamard test method and the procedure is as follows. 

We use the block encoding of $xx^T$ to apply to the state $\ket{\bf 0}_A \ket{\alpha}$ (where $A$ refers to the ancilla system for the block encoding), resulting in the following:
\begin{align}
   \ket{\psi} = \ket{\bf 0}_A x x^T \ket{\alpha} + \ket{\phi_\perp},
\end{align}
where again $ \ket{\bf 0}_A \bra{\bf 0}_A \otimes I \cdot \ket{\phi_{\perp}} = 0$. 

Then we generate $\ket{\bf 0}_A \ket{\alpha}$ in another register. We use $\ket{\psi}$ and $\ket{\bf 0}_A$ in the Hadamard test and observe that their overlap is $\beta$. One may wonder what if the value of $\beta$ is small, e.g., of the order $1/n$. If that is the case, then $1/\beta$ is very large, resulting in a substantial running time. Since the state $\ket{\alpha}$ can be arbitrary, one can heuristically choose an arbitrary unitary $U$ and execute the algorithm. As the value $\beta$ can be evaluated via the Hadamard test, we can first test to see if $\beta$ is large enough. Then the value of $\gamma$ in Lemma~\ref{lemma: 11} is bounded, as claimed. Note that since at each iteration step $t$, we might use a different $U$ and hence the value of $\beta$ at each step is different, the actual value of $\gamma$ in Lemma~\ref{lemma: 11} is the maximum among all values of $\gamma$, hence justifying the running time to be $\mathcal{O}(T \gamma^{2p-2})$. 

The above procedure is quite heuristic as we choose $U$ that provides a non-small value of $\beta$. This is quite practical as it only requires changing the circuit $U$ and using the Hadamard test. Now we provide a more careful analysis to show that, for a given $U$, e.g., one might choose $U$ to be the circuit that generates the initial state $\ket{x_0}$, if one chooses the step size $\eta$ to be sufficiently small, then the value of $\beta$ is guaranteed to be lower bounded. 

We observe that, once we begin with $\ket{x_0} \bra{x_0}$ (or $x_0 x_0^T$ for simplicity), at step $t$, we obtain the new operator
$X_t = x_t x_t^T$
and the new operator in the next iteration is:
\begin{align}
     X_{t+1} = x_{t+1} x_{t+1}^T = \frac{1}{4}  \Big( (I-D(X_t)) X_t (I- D(X_t) )^T   \Big).
\end{align}

As the norm of $D \leq p$ for any input, we have that:
\begin{align}
   | x_0^T x_{t+1} | &= | \frac{1}{2} x_0^T  (I- \eta D(X_t) ) x_t | \\
   &= | \frac{1}{2} x_0^T x_t - \frac{1}{2} \eta x_0^T D(X_t) x_t| \\
   & \geq |   \frac{1}{2} x_0^T  x_t - \frac{1}{2}\eta p x_0^T x_t       | \\
   &= | ( \frac{1-\eta p}{2} ) \cdot x_0^T x_t  |.
\end{align}

By a simple inductive procedure for a total of $(t+1$) iterations, and using the fact that $|x_0|^2 =1/4^t$ as shown previously, we can infer that:
\begin{align}
    | x_0^T x_{t+1} |  \geq \Big( \frac{1-\eta p} {2}\Big)^{t+1} \frac{1}{4^t} = 4 \Big(\frac{1-\eta p} {8}\Big)^{t+1}.
\end{align}
If we choose the unitary $U$ to be the one that generates $x_0$, we have that at the $t$-th step:
\begin{align}
    \beta = x_0^T x_t x_t^T  x_0 = 16 \Big(\frac{1-\eta p} {8}\Big)^{2^t}.
\end{align}
If we want $\beta$ to be lower bounded, for example, greater than $1/4$, then we require:
\begin{align}
    \Big( \frac{1-\eta p} {8}\Big)^{2t} \geq \frac{1}{64}.
\end{align}
The above condition is equivalent to:
\begin{align}
    \eta \leq \frac{1}{p}\Big( 1- 8 \Big(\frac{1}{8}\Big)^{1/t}  \Big),
\end{align}
which shows that if $\eta$ is sufficiently small then $\beta$ is guaranteed to be lower bounded at time step $t$, which in turn means that $\gamma$ (in Lemma~\ref{lemma: 11}) is upper bounded at the same time step. For a total $T$ steps in the gradient descent algorithm, one simply chooses $\eta$ correspondingly by fixing $t = T$ in the above equation. 

\revise{
\section{Power method iterations}
\label{sec: powermethoditeration}
In this section we provide more details on the number of iterations in the power method. Denote the eigenvectors of $A$ as $\{ \ket{E_1},\ket{E_2},...,\ket{E_n}  \}$ and the corresponding eigenvalues of $A$ as $\{ E_1, E_2,..., E_n \}$. Suppose without loss of generalization that $E_1 > E_2 > ... > E_n$, so $E_1 \equiv \lambda_{\max}$ is the largest eigenvalue of $A$.  
According to Theorem 8.2.1 of Ref.~\cite{golub2013matrix1}, by defining a sequence of quantities:
\begin{align}
    x_k &= A^k x_0\\
    \ket{x_k} &= \frac{x_k}{||x_k||} \\
    \lambda_k &=  \bra{x_k} A \ket{x_k} \\
    cos(\theta_k) &= | \braket{E_1,x_k}  |
\end{align}
Then we have that: 
\begin{align}
    \big|  \lambda_k - \lambda_{\max}    \big| \leq \max_{2\leq i \leq n} | \lambda_{\max} - \lambda_i| \tan^2(\theta_0) \Big| \frac{\lambda_2}{\lambda_1} \Big|^{2k}
\end{align}
Since we assume the eigenvalues of $A$ falling within $(-1,1)$, so $  \max_{2\leq i \leq n} | \lambda_{\max} - \lambda_i|  \leq 2$. If we expect an error of $\delta$, we demand that:
\begin{align}
    2 \tan^2(\theta_0) \Big| \frac{\lambda_2}{\lambda_1} \Big|^{2k} \leq \delta
\end{align}
It implies that: 
\begin{align}
    \frac{2 \tan( \theta_0)^2}{ \delta} &\leq \Big|  \frac{\lambda_1}{\lambda_2}  \Big|^{2k} \\
    \longrightarrow  \log\Big( \frac{2 \tan( \theta_0)^2}{ \delta}\Big)  &\leq 2k \log  \frac{\lambda_1}{\lambda_2} \\
    \longrightarrow \log\Big( \frac{2 \tan( \theta_0)^2}{ \delta}\Big)  \frac{1}{ \log \frac{\lambda_1}{\lambda_2}} &\leq 2k 
\end{align}
Let $\Delta = \lambda_1-\lambda_2$ denotes the gap between two largest eigenvalue. Then we have $ \log \frac{\lambda_1}{\lambda_2} = \log \Big( 1 + \frac{\Delta}{\lambda_2} \Big) < 1 + \frac{\Delta}{\lambda_2}$, so from the above we have:
\begin{align}
     \log\Big( \frac{2 \tan^2(\theta_0) }{ \delta}\Big)  \frac{1}{ \Big(1 + \frac{\Delta}{\lambda_2} \Big)} \leq 2k 
\end{align}
Because $1 + \frac{\Delta}{\lambda_2} > \frac{\Delta}{\lambda_2}$, we have:
\begin{align}
     \frac{1}{ \Big(1 + \frac{\Delta}{\lambda_2} \Big)} &\leq \frac{\lambda_2}{\Delta} \\
     \longrightarrow \log\Big( \frac{2 \tan^2(\theta_0) }{ \delta}\Big)  \frac{1}{ \Big(1 + \frac{\Delta}{\lambda_2} \Big)}  &\leq \log\Big( \frac{2 \tan^2(\theta_0) }{ \delta}\Big)   \frac{\lambda_2}{\Delta}
\end{align}
So as long as the right-hand side of the above equation is less than $2k$, then it is guaranteed that the left-side is also less than $2k$. Thus, we arrive at:
\begin{align}
    \log\Big( \frac{2 \tan^2(\theta_0) }{ \delta}\Big)   \frac{\lambda_2}{\Delta} \leq 2k
\end{align}
Now we consider the term inside the logarithmic, $\tan^2(\theta_0) $. Recall that cosine $\cos(\theta_0) = | \braket{E_1, x_0} |$, which is the overlap between the initial vector and $\ket{E_1}$, the eigenvector corresponding to the largest eigenvalue of $A$. Since $\tan^2(\theta_0) = \frac{\sin^2(\theta_0)}{\cos^2(\theta_0)} \leq \frac{1}{\cos^2(\theta_0)} $, it is sufficient to have:
\begin{align}
   \log\Big( \frac{2 \tan^2(\theta_0) }{ \delta}\Big)   \frac{\lambda_2}{\Delta} \leq  \log\Big( \frac{2 }{ \delta  \cos^2(\theta_0)  }\Big)   \frac{\lambda_2}{\Delta} \leq 2k
\end{align}
Hence, by choosing $k = \mathcal{O}\left(  \frac{1}{\Delta} \log\Big( \frac{2 }{ \delta  \cos^2(\theta_0)  }\Big)  \right)$, then it is guaranteed that the estimated value $ \lambda_k $ is $\delta$-close to the true largest eigenvalue of $A$. Two critical factors presented in this power method is the gap $\Delta$ and $\cos(\theta_0)$ -- which is the overlaps between $\ket{x_0}$ and eigenvector of $A$. Hence, the scaling of power method can varies significantly depending on how good initialization is. If the initial vector $\ket{x_0}$ fortunately turns out to be very closed to $\ket{E_1}$, $\braket{E_1, x_0} =\mathcal{O}(1)$, then $k$ can be very small and do not depend on the dimension $n$. On average, it is reasonable to expect that $\braket{E_1,x_0} = \frac{1}{\sqrt{n }}$, thus the value of $k = \mathcal{O}\Big( \frac{1}{\Delta}\big( \log(n) + \log \frac{1}{\delta}\big) \Big)$, which matches the complexity derived in \cite{friedman1998error}. 
}

\end{document}